\documentclass[structabstract]{aa}  
\usepackage{graphicx}
\usepackage{txfonts}
\begin{document}
   \title{Deep {\it Chandra} observation of the galaxy cluster WARPJ1415.1+3612 at $z$=1: }  
   \subtitle{an evolved cool-core cluster at high redshift}


   \author{J.S. Santos\inst{1},
          P. Tozzi\inst{2,3},
          P. Rosati\inst{4},          
          M. Nonino\inst{2},
          and G. Giovannini \inst{5,6}
	}

   \institute{
	     \inst{1} European Space Astronomy Centre (ESAC)/ESA, Madrid, Spain \\
             \email{jsantos@sciops.esa.int} \\
	     \inst{2} INAF, Osservatorio Astronomico di Trieste, via G.B. Tiepolo 11, 
34131, Trieste, Italy \\
	     \inst{3} INFN, Istituto Nazionale di Fisica Nucleare, Trieste, Italy  \\
             \inst{4} European Southern Observatory, Karl-Schwarzchild Strasse 2, 85748 
Garching, Germany \\
             \inst{5} Dipartimento di Astronomia, via Ranzani 1, 40127 Bologna, Italy \\
              \inst{6} Istituto di Radioastronomia-INAF, via P.Gobetti 101, 40129 Bologna, Italy \\             
             }

   \date{Received ... ; accepted ...}


  \abstract
   {} 
   {Using the deepest (370 ksec) \textit{Chandra} observation of a
     high-redshift galaxy cluster, we perform a detailed
     characterization of the intra-cluster medium (ICM) of
     WARPJ1415.1+3612 at $z$=1.03, particularly its core region. We
     also explore the connection between the ICM core properties and
     the radio/optical properties of the brightest cluster galaxy
     (BCG).}
{We perform a spatially resolved analysis of the ICM to obtain
  temperature, metallicity and surface brightness profiles over the
 8--400 kpc radial range.  We measure the following cool-core
  diagnostics: central temperature drop, central metallicity excess, 
  central cooling time, and central entropy.  Using the
  deprojected temperature and density profiles, we accurately derive
  the cluster hydrostatic mass at different overdensities.  In addition to the
  X-ray data, we use archival radio VLA imaging and optical GMOS
  spectroscopy of the central galaxy to investigate the feedback
  between the central galaxy and the ICM. }
{
  Our spatially resolved spectral analysis shows a significant temperature drop 
  from a maximum of 8.0 keV to a projected core value $T_{c}=4.6\pm 0.4$ keV, 
  and a remarkably high central iron abundance peak, $Z_{Fe,c}$= 3.60$^{+1.50}_{-0.85}
  Z_\odot$, measured within a radius of 12 kpc.
  We measure $M_{500}$=$M(r<R_{500})$=2.4$\pm$0.4 $M_\odot$ and a corresponding 
  gas fraction $f_{gas}$=0.10$\pm$0.02.
  The central cooling time is shorter
  than 0.1 Gyr and the entropy $K_{c}$ is equal to 9.9 keV cm$^{2}$.
  We detect a strong [OII] emission line in the optical spectra of the
  BCG with an equivalent width of -25 \AA, for which we derive a star
  formation rate within the range 2 $-$ 8$~M_\odot yr^{-1}$.  The VLA
  data reveals a central radio source coincident with the BCG with a
  luminosity $L_{1.4GHz}$=2.0$\times$10$^{25}$ W Hz$^{-1}$, and a
  faint one-sided jet-like feature with an extent of $\sim$80 kpc. We
  do not find clear evidence for cavities associated to the radio AGN
  activity.  }
{Our analysis shows that WARPJ1415 has a well developed cool-core 
with ICM properties similar to those found in the local
  Universe. Its properties and the clear sign of feedback activity
  found in the central galaxy in the optical and radio bands, show
  that feedback processes are already established at $z\sim 1$ (a
  lookback time of 7.8 Gyr). In addition, the presence of a strong
  metallicity peak shows that the central regions have been promptly
  enriched by star formation processes in the central galaxy already
  at $z>1$. Our results significantly constrain the timescale for the
  formation and self-stabilization of cool-cores.  }

   \authorrunning{J.S.Santos et al.}
   \titlerunning{Deep {\it Chandra} observation of WARPJ1415 at $z$=1}

   \maketitle

%

\section{Introduction}

Galaxy clusters are dynamical environments hosting complex astrophysical phenomena, 
that provide us with a wealth of information on the intricate processes that shape the 
cosmic large-scale structure and galaxy evolution (see reviews by Fabian \cite{fabian94}, 
Rosati, Borgani \& Norman \cite{rosati02}, Voit \cite{voitrev}). 

The intracluster medium (ICM) is the dominant baryonic component of galaxy clusters, a
hot plasma emitting X-ray radiation via thermal bremsstrahlung. 
High-resolution X-ray observations show that the surface brightness of the ICM of 
about half of the local clusters is
peaked in their central regions, where the inferred cooling time of the gas is shorter 
than the typical dynamical time (see Fabian \cite{fabian94} for a review on cooling flows and 
Hudson et al. \cite{hudson} for more recent results). 
However, X-ray spectroscopy always shows a gas temperature floor, indicating that some distributed source 
of heating must stop the cooling process (see Peterson \& Fabian \cite{peterson06} and references therein). 
In the current framework, the dominant heating source that prevents the overcooling of the ICM is likely to be an AGN fueling a massive central black hole, leading to an outburst 
which can heat the ICM via shocks, buoyantly rising bubbles inflated by radio lobes, or dissipation of 
sound waves (see McNamara \& Nulsen \cite{mcnamara} for a review).

The so-called cool-core phenomenon is observed in different wavebands depending on the
cluster component under investigation. 
Among them: the ICM (X-ray), the brightest cluster galaxy (BCG, optical),
cold molecular gas (IR), and the central AGN (radio).
The onset of star formation activity in the central galaxy (otherwise a passive, early-type galaxy)
 is a well-known manifestation of this phenomenon (Crawford et al. \cite{crawford}, McNamara et al. \cite{mcnamara}, Donahue et al. \cite{donahue}). 
However, nuclear activity in the central cluster galaxy is expected to play the major role in 
regulating the cool-core thermodynamics, as suggested by the observed interactions between the 
radio jets and the ICM and the scaling relations among total radio power and ICM properties 
observed in the X-ray band (Birzan et al. \cite{birzan08}). These processes are shown in the spectacular
combined images of nearby clusters in the radio and X-ray bands, where cavities and ripples in the
ICM are observed to correspond to non-thermal radio emission (Blanton et al. \cite{blanton01}; Birzan et al. 
\cite{birzan04}; Wise et al. \cite{wise}; Sanders \& Fabian \cite{sanders07}; 
Sanders, Fabian \& Taylor \cite{sft}, and many others). 

There is increasing evidence
from radio and X-ray observations of local clusters that the
formation of bubbles due to radio jets associated to the central AGN
may effectively satisfy the energy balance between cooling and heating
in cool-core regions (Blanton \cite{blanton01}, Ehlert et al. \cite{ehlert}). In this respect, the radio 
luminosity of the central galaxy provides an important link
between the black hole activity and the state of the intracluster medium. 

To date, systematic studies of the cool-core phenomenon including X-ray, optical and radio 
data to understand how the feedback mechanism shapes the evolution of cool-cores have been 
limited to samples of nearby clusters (e.g., Heckman et al. \cite{heckman}, Birzan et al. \cite{birzan04}, 
Cavagnolo et al. \cite{cavagnolo}, O'Dea et al. \cite{odea}, Mittal et al. \cite{mittal}, 
Sun \cite{sun}, McDonald et al. \cite{mcdonald}), reaching a redshift up to $<z>$=0.3 
(Samuele et al. \cite{samuele}).
These studies show clear correlations involving the radio luminosity of the central AGN, optical emission 
lines (e.g. H$\alpha$), excess IR/UV emission, and the core entropy of the ICM.

The extension of these studies to high-redshift can provide important
clues also on the timescales of the metal enrichment mechanisms of the ICM,
a process that encompasses different phases of the cool-core
phenomenon.  Iron, the main metal locked in the ICM, is produced
primarily by type Ia supernovae, while iron and other heavy elements produced
by both type Ia and type II SNe hosted by the cluster early-type
galaxies (Renzini et al. \cite{renzini}) are eventually diffused
into the ICM via galactic winds and ram pressure striping.  According
to local studies, the excess iron mass found in cool-cores is directly
linked to the brightest cluster galaxies (B\"ohringer et
al. \cite{boehringer}, De Grandi et al. \cite{degrandi04}).
A complete understanding of the formation and evolution of cool-cores should encompass all these aspects.

Recent results indicate a moderate evolution in the bulk of the cool-core cluster population 
out to $z$=1.3, with an apparent lack of very strong cool-core clusters at high redshifts 
(Santos et al. \cite{joana08}; \cite{joana10}). However, to adequately compare the 
characteristics of distant and nearby 
cool-core clusters, we require deep, high-resolution X-ray data of distant clusters. 
The study of cool-cores at high-redshift will not only place 
an upper limit on the formation timescale of cool-cores, but will also reveal the strength of 
the interactions between the diffuse baryons and the brightest central galaxy (BCG) 
at an epoch when the most massive clusters are still assembling.

In this paper we provide a detailed investigation of the ICM of the X-ray selected cluster 
WARPJ1415.1+3612 (hereafter WARPJ1415) at $z$=1.03 (Perlman et al. \cite{perlman}), 
using a recently acquired deep \textit{Chandra} observation.
This cluster was chosen as the strongest high-$z$ cool-core cluster based on 
the analysis of its surface brightness properties (Santos et al. \cite{joana10}). 
Using a multi-wavelength dataset comprising radio VLA data,  archival optical 
GEMINI-GMOS spectroscopy and imaging from HST/ACS F775W, and SUBARU-\textit{Suprime} (BVRiz)
 we study the central galaxy and its interactions with the hot plasma. 
The analysis presented here is the first detailed study of the feedback process in a cool-core at 
such high-redshift. 

The paper is organized as follows: 
in \S 2 we describe the X-ray data reduction procedures, including the detailed 
spatially resolved spectroscopic analysis of the ICM.
In \S 3 we obtain the temperature and metallicity profiles. We perform the deprojection 
analysis and present the mass profile in \S 4.
We investigate the surface brightness properties of the ICM in \S 5
and we accurately measure several cool-core diagnostics, namely $c_{SB}$, the 
central cooling time and central entropy.
In \S 6 we present the radio and optical properties of the BCG and explore the connection 
between the BCG and the ICM core properties. Our conclusions are summarized in \S 7.  

The cosmological parameters used throughout the paper are: $H_{0}$=70 km/s/Mpc,
$\Omega_{\Lambda}$=0.7 and $\Omega_{\rm m}$=0.3. Quoted errors are at the 1-$\sigma$ 
level, unless otherwise stated.

\section{X-ray data reduction and spectral analysis}

\subsection{The X-ray dataset}

The galaxy cluster WARPJ1415 was detected in the Wide Angle ROSAT
Pointed Survey (Jones et al. \cite{jones98}, Perlman et al. \cite{perlman}). This survey 
provided several high-redshift galaxy clusters based on
serendipitous detections of extended sources in targeted ROSAT PSPC
observations. The survey covers an area of 71 deg$^{2}$ and contains a
complete sample of 129 X-ray selected clusters down to a flux limit of
$S \sim 6.5 \times 10^{-14}$ erg s$^{-1}$ cm$^{2}$.

A deep, 280 ksec ACIS-S observation of WARPJ1415 was awarded to our
group in \textit{Chandra} AO 12 (PI J. Santos). An additional 90 ksec 
observation with ACIS-I (taken in 2004, PI H. Ebeling) was used in
our analysis.  The \textit{Chandra} exposure time sums up to a total 370
ksec ({\tt obsid} 4163, 12255, 12256, 13118, 13119).  
This dataset represents the deepest \textit{Chandra} observation of a
cluster at $z\simeq 1$, enabling the most detailed X-ray analysis of
any such high redshift galaxy cluster.  Previous studies of distant
X-ray clusters with comparable depth used both shallow \textit{Chandra} and XMM-Newton
data (see Maughan et al. \cite{maughan08}), and therefore do not reach the angular
resolution to study the inner regions of the ICM distribution.  In the
case of WARP1415, angular resolution has a paramount relevance, since
it allows us to resolve the density, temperature and metal distribution within the 
cool-core region on a scale as small as $\sim$10 kpc, and hence to shed
light on the complex physics of the cool-core region at a look back
time of 8 Gyr.

We also examined the XMM-Newton archival data on WARPJ1415. We find
that after removing the time intervals of high-background, the useful
exposure time is about 17 ks for MOS and 14 ks for PN. Given the
modest exposure time and the much lower angular resolution of XMM-Newton, we
do not include this dataset in our analysis.

\subsection{Data reduction}

We performed a standard data reduction starting from the level=1 event
files, using the {\tt CIAO 4.3} software package, with the most recent
version of the Chandra Calibration Database at the time of writing
({\tt CALDB 4.4.5}).  We ran the {\tt acis$\_$process$\_$events} task
to update the data calibration, including the {\tt tgain}
correction\footnote{See http://cxc.harvard.edu/ciao/why/acistgain.html}.  Since both ACIS-I
and ACIS-S observations were taken in the VFAINT mode, we set the {\tt
  check$\_$vf$\_$pha} key to {\tt yes} to flag background events that
are most likely associated with cosmic rays and distinguish them from
real X-ray events. With this procedure, the ACIS particle background
can be significantly reduced compared to the standard grade selection.
We also applied the CTI correction to the ACIS-I observation (taken with
the temperature of the Focal Plane equal to 153 K).  This procedure
allows us to recover the original spectral resolution partially lost
because of the CTI (see Grant et al. \cite{grant}).  The correction applies
only to ACIS-I data, since the ACIS-S3 did not suffer from radiation
damage.  The data were filtered to include only the standard event
grades 0, 2, 3, 4 and 6.  We did not remove the {\tt
  acis$\_$detect$\_$afterglow} correction since all our data have a
SDP version larger than 7.4.0\footnote{see http://cxc.harvard.edu/ciao/threads/acisdetectafterglow/}.
Nevertheless, we checked visually for hot columns left after the
standard reduction, finding none.  We also identified the flickering
pixels as the pixels with more than two events contiguous in time,
where a single time interval was set to 3.3~s. For exposures taken in
VFAINT mode, there are no flickering pixels left after filtering out
bad events.  We filtered time intervals with high background by
performing a $3\sigma$ clipping of the background level in the 0.5$-$7
keV band by using the script {\tt analyze\_ltcrv}.  We did not detect
any high background interval in both ACIS-I and ACIS-S data (nominally
the removed exposure time is less than 0.5\% for ACIS-I and less than
0.2\% for ACIS-S).  We remark that our spectral analysis is not
affected by any possible residual flare unnoticed in the 0.5-7 keV
band, since we compute the background from source-free regions around
the cluster, thus taking into account any possible spectral distortion
of the background itself. Finally, all the ACIS-S {\tt obsid} were 
merged together with the {\tt merge\_all} tool. ACIS-S and ACIS-I are kept 
separated both for imaging and for spectral analysis, as described in the following sections.

\subsection{Spectral analysis}

We produced soft (0.5-2 keV) and hard (2-7 keV) band images with the full
\textit{Chandra} resolution (1 pixel = 0.492 arcsec).  We computed the cluster
center in the soft-band by measuring the position at which the
signal-to-noise ratio (S/N) within circles of different radii is maximum.  
Following this procedure we assign the X-ray center of the cluster to 
RA=14:15:11.08, DEC=+36:12:03.1.  
This position also corresponds to the peak of the surface brightness emission.  
We remark that we did not detect any point source which could significantly 
contribute to the core emission, by carefully inspecting the 0.5-2.0 keV and the 2.0-7.0 keV X-ray images.

The spectral analysis is performed as follows: starting from the
cluster center, we draw rings at different radii and compute the S/N
in the total 0.5-7 keV band for the ACIS-S image, which has a signal
much stronger than the ACIS-I image.  In order to achieve a good
compromise between the S/N ratio in each ring and the total number of
rings, we set a minimum S/N $\ge 24$ within 25$\arcsec$, decreasing at
larger radii down to 14 in the outermost radius.  This ensures that we
have always more than 650 net counts in the 0.5-7 keV band in each
ring (with the exception of the central bin with 500 net counts), for
a total of nine rings.  We adopted the same set of circular rings in
the ACIS-I image, despite the net counts in each ring are typically
less then 100.  In the last radial bin (300-400 kpc), we used only
ACIS-S, since there is no useful signal in the ACIS-I data.  

We extracted spectra for each ring from the ACIS-S and ACIS-I data separately.  
The response matrices and the ancillary response matrices were computed 
for each {\tt obsid} respectively with {\tt mkacisrmf} and {\tt mkwarf}, for the same regions where the 
spectra were extracted.  A single {\tt arf} and {\tt rmf} files for each ACIS-S spectrum were 
finally obtained by summing over the 4 ACIS-S {\tt obsid} with {\tt addarf} and {\tt addrmf}, respectively.
 We detected 7500 (6200 in ACIS-S and 1300 in ACIS-I) photons within 50$\arcsec$, 
 corresponding to 400 kpc.

We selected the background from empty regions of the same CCD in 
which the cluster is located. This is possible since the signal from the
cluster has a total extension of less than 1 arcmin, as opposed to the
8$\arcmin$ size of the ACIS-I/-S chips. The background region is scaled
to the source files by the ratio of their  geometrical areas.  In principle,
the background regions may partially overlap with the outer virialized
regions of the clusters. However, the cluster emission from these
regions is negligible with respect to the instrumental background, and
does not affect our results.  Our background subtraction procedure, on
the other hand, has the advantage of providing the best estimate of
the background for that specific observation.

The background is extracted (both for ACIS-I and ACIS-S data) from 3 circular regions 
nearby the clusters, up to a maximum distance of 3 arcmin, and with a minimum distance of 1.2 arcmin. 
The size of each background region varies between 50 to 70 arcsec in radius, in order to ensure 
the background is sampled from a region larger by at least a factor of 10 with respect to the
outermost ring. We repeated the measurement with other background regions selected with these 
criteria and found no significant differences in the final results.

The background spectra were scaled by the geometrical factor between the background and source regions before subtraction.
Our spectral analysis is expected to be robust against background variations and vignetting effects thanks to the high S/N set for our circular rings (S/N of 25 except in the two outermost rings, where we have S/N=20 and 15) and to the sparse sampling of the background.  To show this, we also compute the scaling factors using the ratio of the exposure maps of the source and background regions, and repeat the spectral analysis.  In all cases we obtain results in excellent  agreement with those obtained with the simple geometrical scaling.  We remark that a full treatment of the vignetting effects would imply modeling the background by treating each component separately.  We do not apply this procedure in our paper since it would not introduce any noticeable improvement in the final results.

The spectra of each ring were analyzed with XSPEC v12.6 (Arnaud et
al. \cite{arnaud}) and fitted with a single-temperature {\tt mekal}
model (Kaastra \cite{kaastra}; Liedahl et al. \cite{lied}) using the
solar abundance of Asplund et al. (\cite{asplund})\footnote{The values
  normalized to Asplund et al. (2005) are a factor 1.6 larger than the
  ones normalized to Anders \& Grevesse (1989) which have been
  extensively used so far in the literature.}.  The fits were performed
over the energy range $0.5-8.0$ keV.  The free parameters in our
spectral fits are temperature, metallicity and normalization.  The
local absorption is fixed to the Galactic neutral hydrogen column
density ($N_H$) measured at the cluster position (Wilms et
al. \cite{wilms}) and equal to $N_H=1.05 \times 10^{20}$ cm$^{-2}$,
and the redshift is fixed to $z=1.03$.  We used Cash statistics
applied to the source plus background, which is preferable for low S/N
spectra (Nousek \& Shue \cite{nousek}).

The spectra were binned only to one photon per energy bin.  The use of Cash statistics 
with background subtracted spectra has been used in previous works, and validated with 
spectral simulations in the case of typical AGN (see Figure 2 in Tozzi et al. \cite{tozzi}).  
The same results were obtained for thermal spectra typical of clusters.

\begin{figure}[h]
\begin{center}
\includegraphics[width=6.cm,angle=-90]{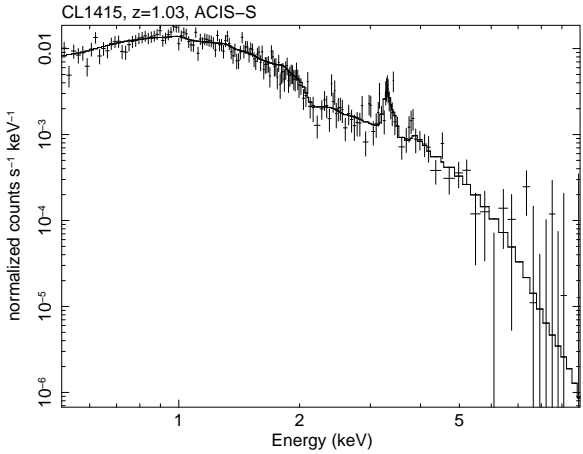}
\end{center}
\caption{Folded global spectrum of WARPJ1415 extracted from a circular region
  with radius of 32$\arcsec$ in the ACIS-S data (crosses) fitted with a {\tt
    mekal} model (continuous line). The redshifted $K_{\alpha}$ line complex of 
    Hydrogen-like and Helium like Iron (Fe XXV, XXVI) is clearly visible at $ \sim 3.5$ keV.} 
 \label{maxsn}
\end{figure}

The global temperature and abundance values were measured in the region
of maximum S/N in the ACIS-S data, which corresponds to a radius
$R_{ext} = 32\arcsec$ centered around the cluster position. 
The spectrum extracted from this region allows us to measure temperature and Fe 
abundance with unprecedented accuracy at such an high redshift. 
The best fit values are $kT = 6.82^{+0.43}_{-0.34}$ keV and $Z =
0.88^{+0.11}_{-0.10} Z_\odot$. 
In addition, the strong $K_\alpha$ Fe line allow us to measure the redshift with an 
accuracy of less than 1\%: $z_{fit} = 1.028_{-0.007}^{+ 0.008}$ (see Fig. \ref{maxsn}).
The same analysis performed with the ACIS-I
data provides consistent results with larger uncertainties. These
global values are still consistent with the values measured with a
previous CALDB version by Balestra et al. (\cite{balestra}), using the 90 ksec
ACIS-I data.
Our temperature measurement is in agreement, albeit with significantly improved accuracy,
 with those found by Maughan et al. (\cite{maughan06})
in their analysis of the XMM-Newton data, and by Maughan et al. (\cite{maughan08}) in their analysis 
of the ACIS-I \textit{Chandra} data. However, we note that previous measurements of the Fe abundance with 
\textit{Chandra} gave very low values, at variance with our findings.

We will not use the average values of temperature and metal abundance obtained from the highest S/N region in the rest of this study.

\begin{figure*}
\begin{center}
\hspace{-1.5cm}
\includegraphics[height=7.5cm,angle=0]{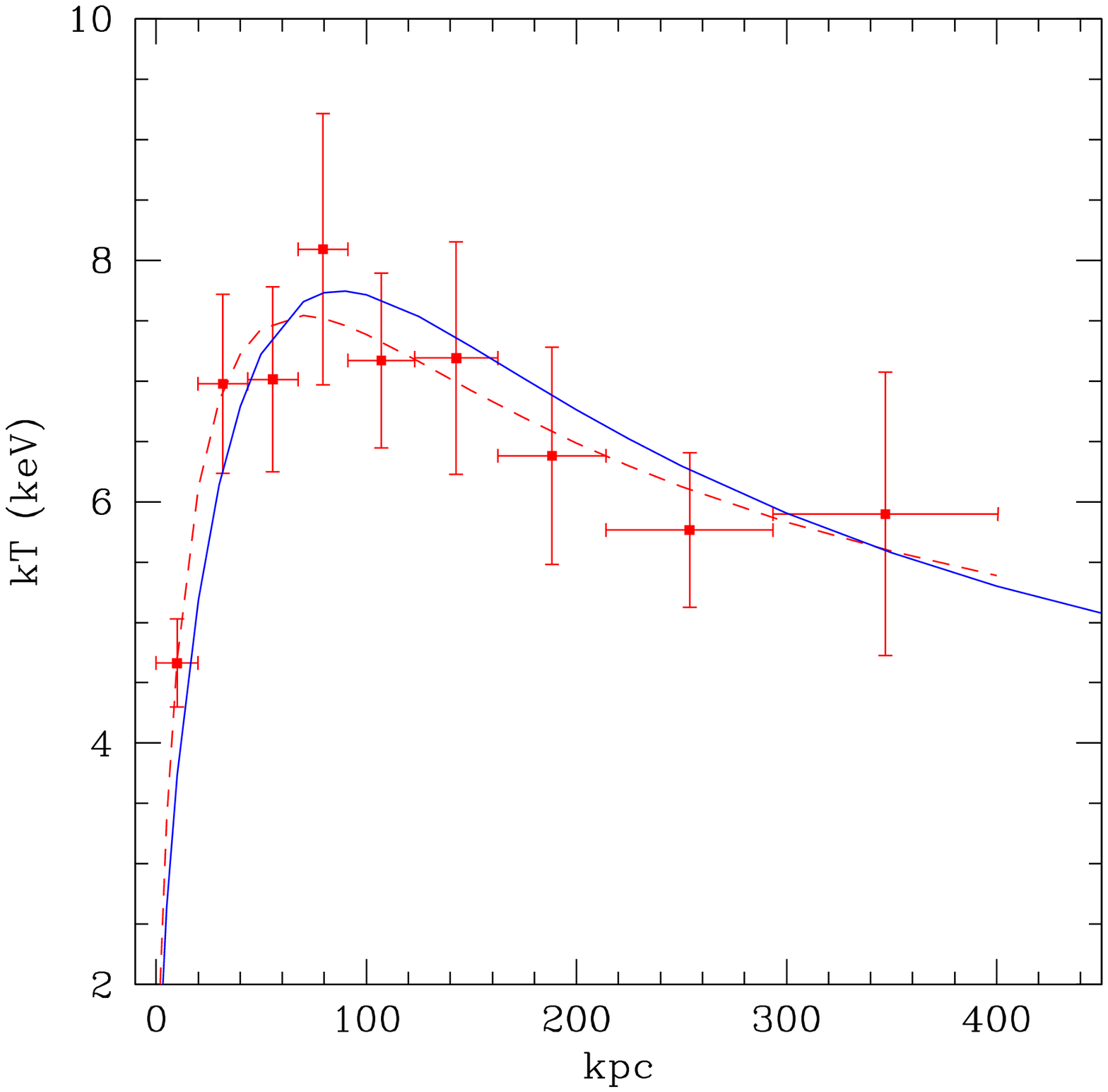}
\includegraphics[height=7.5cm,angle=0]{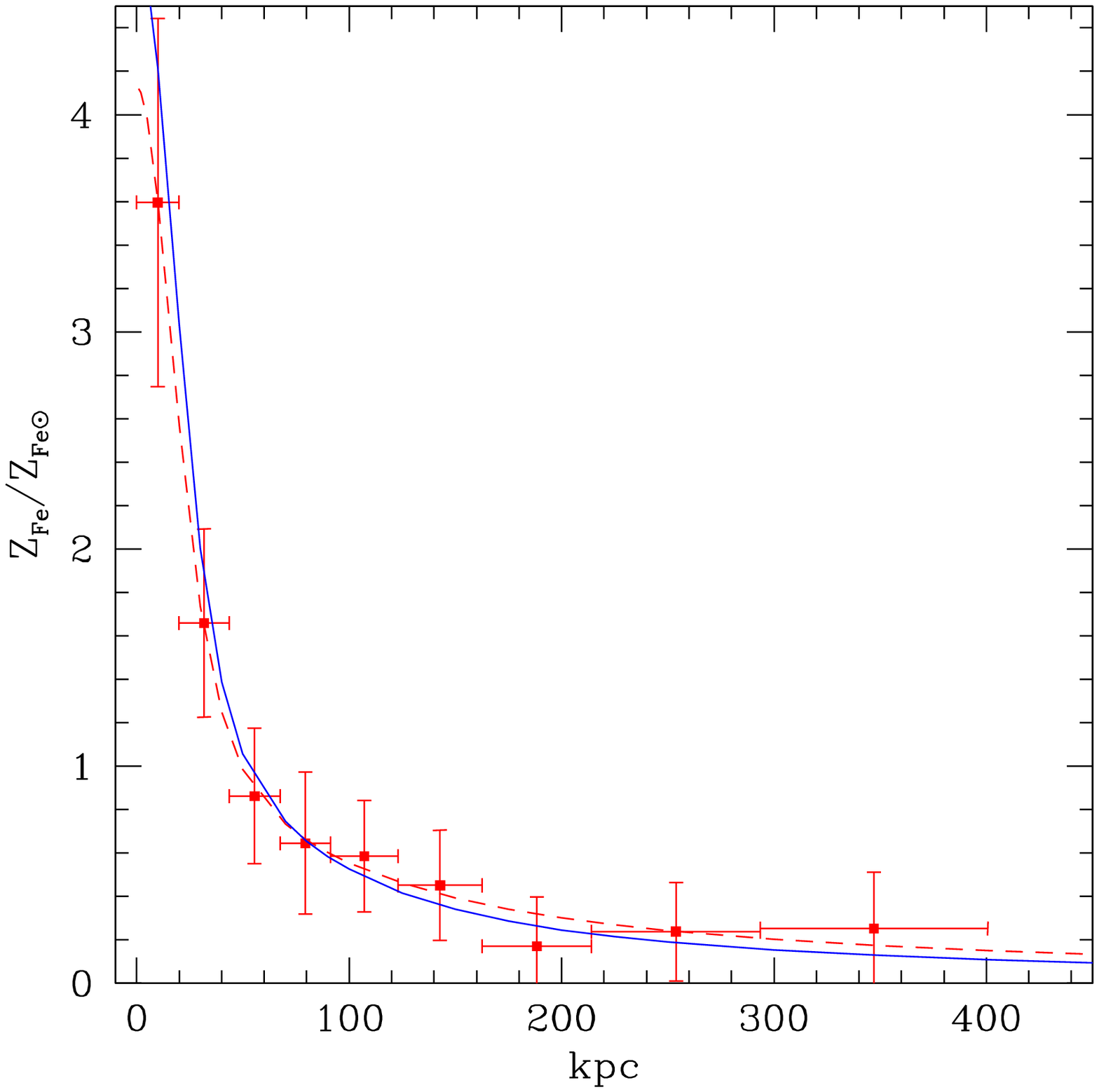}
\end{center}
\vspace{-0.7cm}
\caption{{\sl Left panel:}  Observed projected temperature profile of WARP1415 (red points) 
and the corresponding fit (red dashed line) compared to the deprojected measured temperature 
values (blue solid line).  {\sl Right panel:} same as
  left panel for the Iron abundance profile. The 68\% uncertainty in the fitted curves is very close to the typical error bars of the measured projected quantities at similar radii, and it is not shown for the sake of clarity.} 
 \label{tdeproj}
\end{figure*}

\section{Projected temperature and metallicity profiles}

Nearby cool-core clusters are characterized by several particular
features in the central ICM, most of which require spatially
resolved X-ray data and are therefore challenging to measure in
the typical low S/N observations of high-redshift clusters. 
In the following subsections, we discuss in detail the spatially
resolved properties of the ICM in WARPJ1415. 

\subsection{Projected temperature profile and map}

The most distinctive sign of a cool-core is a temperature drop towards
the cluster center, usually down to a third/half of the global cluster
temperature (see, e.g., Peterson et al. \cite{peterson03}).  A resolved temperature
drop at high-redshift has only been observed in 3C186 at $z$=1.1
(Siemiginowska et al. \cite{aneta}), a cluster that harbors a bright quasar in
its center.  More recently, the analysis of relatively shallow
\textit{Chandra} data of SPT-CL J2106-5844, a galaxy cluster at
$z$=1.1 discovered with the South Pole Telescope, hints at a
temperature drop from 11.0$^{+2.6}_{-1.9}$ keV to 6.5$^{+1.7}_{-1.1}$ keV 
in the core (Foley et al. \cite{foley}), with a drop factor of $1.7 \pm 0.5$.  
Similarly, the very massive merging cluster ACT-CL J0102-4215  at $z$=0.87, 
shows temperature variations ranging from 22$\pm$6 to 6.6$\pm$0.7 keV (Menanteau et al. \cite{menanteau}), 
corresponding to a drop factor of $\sim$2, from the global, core-excluded temperature, to the cluster center.

To measure the projected temperature profile of CL1415 we fitted
the spectra of the rings obtained with the procedure described
in \S 2.  For each ring we combined the ACIS-S and ACIS-I data.
Despite the fact that the majority of the signal is provided by the
ACIS-S data, the use of the ACIS-I spectra allows us to obtain slightly
smaller error bars with respect to the analysis based on the ACIS-S
data only.  In addition, the ACIS-S analysis shows no significant
difference with respect to the combined analysis.  Therefore, from now
on we refer to the combined (ACIS-I+ACIS-S) spectral analysis.  

The projected temperature profile is shown in Figure~\ref{tdeproj},
left panel.  We are able to trace the temperature profile out to 400
kpc with an accuracy of the order of 10\%, reaching 20\% in the outermost
bin (1-$\sigma$ error).  The innermost bin has a size of $\sim 20$ kpc,
while the outermost bin has a size of $\sim 100$ kpc.  We measure a
significant central temperature drop with $T_{c}=4.6\pm 0.4$ at $R <
20$ kpc, which is nearly half of the maximum cluster temperature, $kT
= 8.0 \pm 1.2$ keV, reached at $r= 80$ kpc.  
The ICM temperature drop, from the maximum value to the core value, is measured to 
be $1.7 \pm 0.3$.  Clearly, the observed drop depends also on the capability 
of measuring the temperature in the very center of the cool core, which in turn 
depends on the angular resolution and on the S/N.  Therefore, it is dangerous to 
use solely the measured temperature drop as an indication of the cool core strength.

\begin{figure}
\begin{center} 
\includegraphics[height=6.7cm,angle=0]{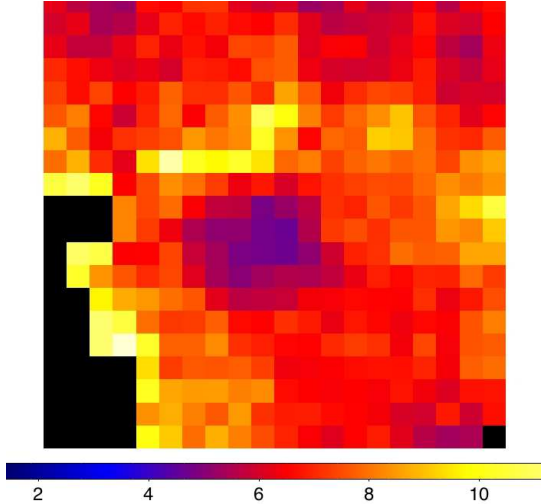}
\end{center}
\vspace{-0.5cm}
 \caption{Temperature map obtained in a square
  region 40$\arcsec \times$40$\arcsec$ centered on the cluster.  The
  colorbar indicates the temperature scale in keV.  Pixels where the 1
  sigma error is larger than 50\% are masked in black.  See text for
  details.}
 \label{tprofile}
\end{figure}

To further investigate the temperature structure of WARP1415, we
produced a temperature map.  We map a square region of 40'' by side
centered on the cluster, extracting circular regions spaced by 2''.
Each circular region has a radius ranging from 3'' to 10'' going from
the cluster center to the outer regions.  Clearly, the spectra assigned
to each pixel are not independent to each other, so the temperature map
is actually a smoothed map, with a smoothing length increasing with
the distance form the cluster center.  The temperature map of WARP1415
is shown in Figure \ref{tprofile}, where we masked out
the pixels where the 1 sigma error on the temperature is larger than
50\%.  In addition to the cool-core, which appears smooth and round
within a radius of 3'' (the minimum smoothing length of the map) we
notice some anisotropy in the temperature profile, with a difference
of about -2 keV in one sector at a distance of about 80 kpc (the
distance where the temperature is maximum).  We will discuss further
this feature in Section 6.

\subsection{Projected Iron abundance profile}

A prominent peak in the Iron distribution is always associated with
the ICM of local cool-core clusters.
The origin of this iron excess with respect to the almost constant
value measured in the outer regions is ascribed mostly to Type Ia
supernovae  (e.g. De Grandi et al. \cite{degrandi04}).  There is no consensus yet in the literature on the origin
of the iron peak and timescale of its buildup.  While some studies of
local cool-cores favor long enrichment times ($>$5 Gyr), and describe
the metal excess as a long lived phenomenon (B\"ohringer  et
al. \cite{boehringer}), the lack of high-$z$ data has prevented a more accurate
assessment of this important aspect.

In this section, we trace the spatial distribution of iron in the
cluster ICM out to $r$=400 kpc, as obtained by the spatially resolved,
combined spectral analysis.  In Fig. \ref{tdeproj}, we present the ICM
Iron abundance profile in solar units, where the solar abundance is
set to the value of Asplund et al. (\cite{asplund}).  We detect an unusually high
central Fe value in the ICM of WARPJ1415, $Z_{Fe,c}/Z_{\sun}=
3.60_{-0.85}^{+1.5}$.  Even taking into account the associated large
error bar, such a high abundance level has only been reported in the
local cluster Centaurus (Graham et al. \cite{graham}).  The implication of this
high Iron concentration and the associated Iron mass are discussed  in
Section 6.5.  We investigated also the Iron abundance map obtained
along with the temperature map, but we do not find any significant
anisotropy in the Iron abundance distribution.

\subsection{Other metals}

In addition to Iron, alpha elements (Si, Ni, S, Mg) produced by
core-collapse supernovae (SNII) contribute significantly to the
enrichment of the ICM (e.g. De Grandi \& Molendi \cite{degrandi09}).  A spatially
resolved analysis of metals other than Iron is not feasible.  Nevertheless,
we searched for emission line of other elements in the spectrum
extracted from the inner $32''$.  
Metal lines were visually explored changing manually the abundance of each element. 
A given element was then removed whenever its presence was irrelevant to the best fit spectrum.
Therefore, all metals were initially unconstrained, and progressively we froze
 to 0.3 the metals which do not contribute significantly
to the folded spectrum.  This procedure can be used to identify
possible low S/N emission lines, by setting the corresponding element to zero 
the C-statistics increases by $\Delta C >3$.  With this method, we confirmed the
detection of the elements Si, S and Ni at 2 $\sigma$ level.  The
nominal best fit values are: $Z_{Si}=1.9 \pm 0.7 ~Z_\odot$;
$Z_{S}=1.3 \pm 0.7 ~Z_\odot$ and $Z_{Ni}=3.6 \pm 1.7 ~Z_\odot$.  
This is the first detection of these elements in the
spectrum of a cluster at $z\simeq 1$, but the lack of spectral
resolution and S/N prevent us from computing a
meaningful $\alpha/Fe$ profile and to derive robust constraints on the
enrichment sources of the ICM.

\section{Deprojection and mass profile}

The next step in our analysis is to compute the deprojected temperature and metal
abundance profiles, and then to compute total and gas masses.
The direct deprojection of the observed profile with the {\tt projct}
model within XSPEC is not feasible due to the errors on the
temperature and the coarse binning.  For the best exploitation of our
data we proceeded as follows.  First we fitted the projected
temperature and Iron abundance profiles.  For the temperature we use the
functional form of Vikhlinin et al. (\cite{vikhlinin06}):
\begin{eqnarray}\label{e.kt}
kT(r) & = & kT_0
\frac{(r/r_{cool})^{a_{cool}}+kT_{min}/kT_0}{(r/r_{cool})^{a_{cool}}+1} 
\frac{(r/r_t)^{-a}}{(1+(r/r_t)^b)^{c/b}}\,\, ,
\end{eqnarray}

\noindent where we set $b=c/0.45$ (as suggested in Maughan et
al. (\cite{maughan07}) and therefore we are left with 7 free parameters.  

This functional form is usually taken to represent the 3D temperature distribution, and projected to be fit to the observed projected temperature profile. Instead, we treat $r$ in Eq. 1 as a projected radius and fit the model to the projected temperature profile directly. The best fitting model is then deprojected to obtain the 3D temperature profile.


As for the Iron abundance, we adopt a double beta model (Cavaliere \& Fusco-Feminano \cite{cavaliere}):

\begin{eqnarray}\label{e.met}
Z/Z_\odot = Z_{in}/Z_\odot \frac{1}{(1+(r/r_1)^{2})^{\beta_1}} + Z_{out}/Z_\odot \frac{1}{(1+(r/r_2)^{2})^{\beta_2}} \,\, ,
\end{eqnarray}

\noindent which has 6 free parameters.
Therefore we can estimate the projected temperature and Fe abundance of WARPJ1415 at any given radius using equations 1 and 2 with the best-fit parameters plugged in.  The next step is to deproject directly the analytical renditions of the temperature and  abundance profile at the same time.  Clearly, when deprojecting the best fit temperature and metallicity profiles, we want to preserve the full information coming from the surface brightness in order to directly obtain also the electron volume density $n_e$ as a function of the radius.  We adopt a finer binning of the surface brightness profile, with the requirement of keeping the error on the normalization of each spectrum at the level of 5-10\%. This is obtained by selecting rings with a minimum S/N of 15.  We used 20 rings with a width ranging from 2\arcsec to 15\arcsec for increasing radii. We assigned a temperature and Iron abundance to each ring according to equations 1 and 2, respectively, computed at the central radius of each ring.  We normalized each spectrum in order to have the same predicted net count rate as observed in the real image in the 0.5-2 keV band.  We then simulated a spectrum for each ring with a very high S/N.  The deprojection was obtained with {\tt projct}, so that we include at the same time the effects of temperature and Fe abundance. In Figure 2, we show the projected data with the best-fit model (dashed line) and the corresponding deprojected profile (solid line).

\begin{figure}[h]
\begin{center}
\hspace{-1.cm}
\includegraphics[height=6.7cm,width=7.8cm,angle=0]{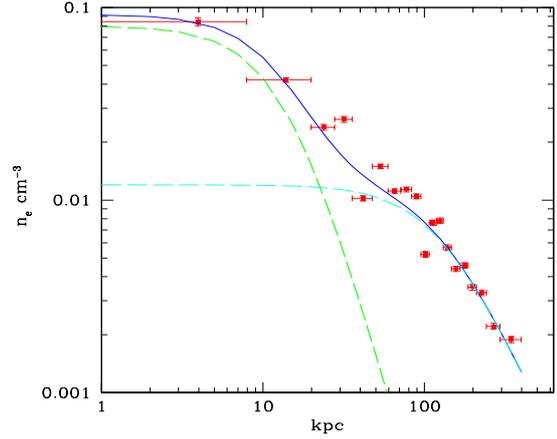}
\end{center}
\vspace{-0.7cm}
\caption{Deprojected electron density profile (red points) and double beta model
  best fit show in solid blue line. The two individual $\beta$ model components are shown in dashed lines.  }
 \label{neprofile}
\end{figure}

The best fit normalization of the deprojected spectra is linked to the electron density 
$n_{e}$ through the relation
\begin{equation}
Norm = {{10^{-14}}\over {4\pi (D_a (1+z))^{2}}} \int n_e n_H dV
\end{equation}

\noindent where $D_a$ is the angular diameter distance to the source (cm), and $n_e$ 
and $n_H$ (cm$^{-3}$)  are the electron and hydrogen densities, respectively.
We fitted the deprojected $n_e$ profile with a double beta-model.  The electron
density profile and the double beta model best fit are shown in Figure
\ref{neprofile}.  
The deprojected profiles of the temperature and of the
electron density will be used to compute the total hydrostatic mass, while the
deprojected Iron abundance profile will be used to measure the Iron
mass (see Section 6.5).  In this section we only compute the total
dynamical mass up to $\sim 400$ kpc.

To measure the cluster mass we adopt the usual spherical symmetry and
hydrostatic equilibrium assumption, which leads to the simple equation (Sarazin \cite{sarazin}):

\begin{equation}
M(r) = - 4.0 \times 10^{13} M_\odot T\,{\rm(keV)} \, r\, {\rm(Mpc)} \left ( \frac{d log \,
  n_{e} } {d log \, r} + \frac{d log \,T } {d log \, r} \right )  .
\end{equation} 
 \label{mass}

\begin{figure}[h]
\begin{center}
\hspace{-1.5cm}
\includegraphics[height=6.7cm,width=7.8cm,angle=0]{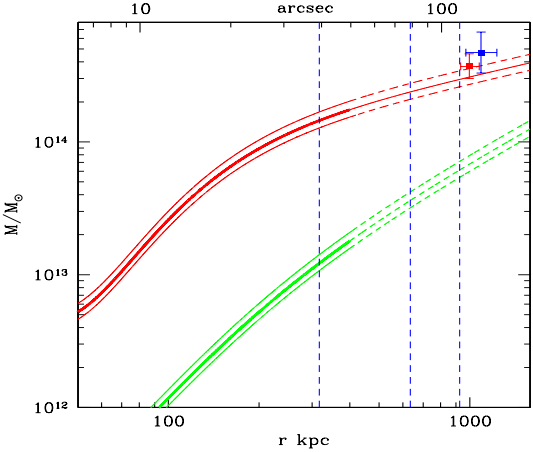}
\end{center}
\vspace{-0.7cm}
\caption{Total mass for WARPJ1415 from hydrodynamical equilibrium
  (red continuous line) and gas mass (green dashed line).  The two dashed lines
  above and below the total mass profile show the 1 $\sigma$
  confidence interval.  The profiles are shown with thick lines up to 400 kpc. 
  Thin lines at $r>400$ kpc show the extrapolation beyond the X-ray detectable emission.  
  Dashed vertical lines
  corresponds to $R_{2500}$, $R_{500}$ and $R_{200}$, from left to right. The red data point 
  refers to the mass derived with the global cluster temperature, whereas the blue point corresponds to 
  the weak lensing mass (Jee et al. \cite{jee}). }
 \label{masstot}
\end{figure}

The logarithmic derivative of the gas density and temperature
profile were computed numerically from our best-fit
profiles.  When we compute the total dynamical mass profile, we also
compute the average density contrast with respect to the critical
density at $z=1.03$, so that we can solve the equation
$M_{\Delta}(r_{\Delta})=\Delta4/3\pi r_{\Delta}^{3} \rho_{c}(z_{cl})$
to measure the radius where the average density level is $\Delta$.
Typically, mass measurements are reported for $\Delta=2500, 500, 200$.  Our
results are summarized in Table \ref{masstab}.  The total mass profile
is shown in Figure \ref{masstot} (solid line) with the corresponding 68\% uncertainty levels.  
The 1 $\sigma$ confidence intervals on the mass are computed by assuming 
that the relative errors on the deprojected temperature profile are equal to those on the 
projected values.  Similarly, the relative error on the electron density is equal to half the error 
on the spectral normalization (therefore at the level of 5\% as described above).  This is a consequence 
of the fact that we do not deproject directly the observed profiles, but rather the analytical fitting formulae, 
therefore we do not introduce additional noise.  Clearly the procedure is correct as far as the fitting formulae accurately reproduce the projected profiles.  This is a fair assumption, given the large number of free 
parameters, which allow us to neglect systematic errors associated to the assumed analytical models.  

\begin{table}
 \centering
\caption{Total cluster mass and gas fraction measured at three radii corresponding to the
  overdensities of 2500, 500 and 200 with respect to the critical density at redshift $z=1.03$.}
\label{masstab}
\begin{tabular}{@{}ccccccccc@{}}\hline
$\Delta$ & $R_\Delta$ (kpc) & $M_{tot}/M_\odot$ & $ f_{gas}$   &  \\ \hline 
2500  &  $317^{+22}_{-18}$ & $1.49_{-0.24}^{+0.33} \times 10^{14}$ & 0.084$\pm$0.016 \\ 
500   &  $635_{-33}^{+41}$ & $2.40_{-0.36}^{+0.44} \times 10^{14}$ & 0.10$\pm$0.02 \\ 
200   &  $926_{-47}^{+57}$ & $3.0_{-0.4}^{+0.6} \times 10^{14}$ &  0.20$\pm$0.03 \\
\hline
\end{tabular}
\end{table}

We remark that we measured temperature and density up to 400 kpc, which
is the upper bound of the last bin with significant signal.  Beyond
this radius, all the derived quantities are obtained by extrapolating
the best-fit profiles.  As an example, if we assume $kT=const$ for
$R>400$ kpc, instead of extrapolating the analytical temperature
profiles, we measure $M_{200} = 3.7_{0.6}^{+0.7} \times 10^{14}
M_\odot$ (a value 25\% higher).  The weak lensing analysis of
WARPSJ1415 has been recently published in Jee et al. (\cite{jee}) using 
HST/ACS data, yielding a total mass $M_{tot}(r<1.09 $ Mpc$)=4.7^{+2.0}_{-1.4} \times
10^{14} M_\odot$, that is consistent with the X-ray hydrostatic
$M_{200}$ for $kT=const$ above 400 kpc, and marginally consistent with
that obtained with a straight extrapolation.  We defer to a
forthcoming paper the detailed discussion of
the X-ray mass profile of WARPJ1415, as well as independent
measurements of the total mass at different radii from a dynamical
analysis of all cluster members and the modelling of a strong lensing
system.

We also computed the total baryonic mass contributed by the ICM.  The
mass in baryons is obtained by directly integrating the electron
density shown in Figure \ref{neprofile}, assuming the
$n_{H}$=$n_{e}$/1.2.  The measured gas fraction at $r_{500}$ is
$0.10 \pm 0.02$ (see Table 1 for $f_{b}$ measured at different radii), typical of 
other distant clusters (see e.g. Ettori et al. \cite{ettori}).

\section{Surface Brightness properties, cooling time and entropy}

The most immediate observational signature of the presence of a
cool-core is a central spike in the surface brightness profile of a
cluster. In this section, we use the cluster surface brightness
properties to derive several cool-core diagnostics.

\subsection{Surface Brightness profile}

We computed the azymuthally averaged surface brightness profile out to
$r$=1 Mpc using the vignetted corrected ACIS-S merged image.  
The isothermal $\beta$-model proposed by Cavaliere \&
Fusco-Femiano (\cite{cavaliere}) is often used as a simple description
of the X-ray surface brightness profiles of galaxy clusters.

\begin{table*}
\caption{Single- and double-$\beta$ model fit parameters: (1) central SB in cts/s/arcmin$^2$; (2) 
slope $\beta$1; 
(3) core radius 1 in kpc; (4) reduced $\chi$2; (5) (6) and (7) correspond to the second model component: 
central SB; slope $\beta$ 2 and core radius 2, respectively. Errors are not presented when the parameter value is
at the limit imposed by the fitting procedure.}  

\label{table:2}      
\centering           
\begin{tabular}{l | lllllll} 
\hline                
 {\bf Fit}  &  {\bf S01 }    & {\bf $\beta_{1}$} & {\bf $rc_{1}$}   & {\bf $\chi$2} &  {\bf S02} &  {\bf $\beta_{2}$} & {\bf $rc_{\begin{small}\begin{footnotesize}                                                                                                                                               \end{footnotesize}                                                                                                                                  \end{small}2}$}  \\   
   &  (1)  &  (2)     & (3) & (4)   & (5) &  (6) &  (7)   \\   
\hline                        

1-$\beta$  & 0.135$\pm$0.007  &  0.51  &  50$\pm$2   & 12.27 & - & - & - \\
2-$\beta$  & 0.457$\pm$0.067  &  0.60$\pm$0.15  &  14$\pm$4   & 3.89  & 0.044$\pm$0.011 & 0.74$\pm$0.08 & 150$\pm$27 \\

\hline  
\end{tabular}
\end{table*}

\begin{figure*}
\begin{center}
\includegraphics[width=7.5cm,angle=0]{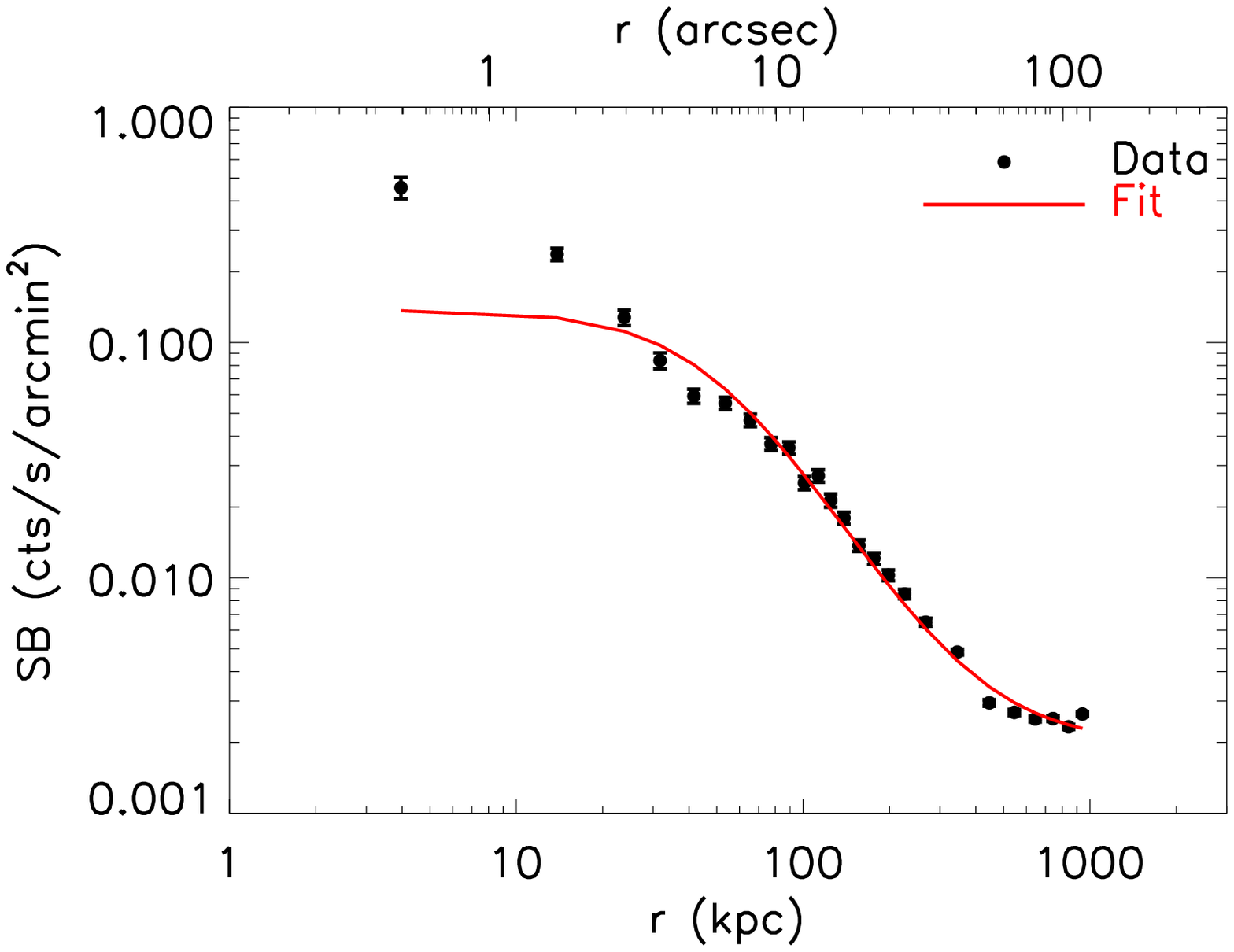}
\includegraphics[width=7.5cm,angle=0]{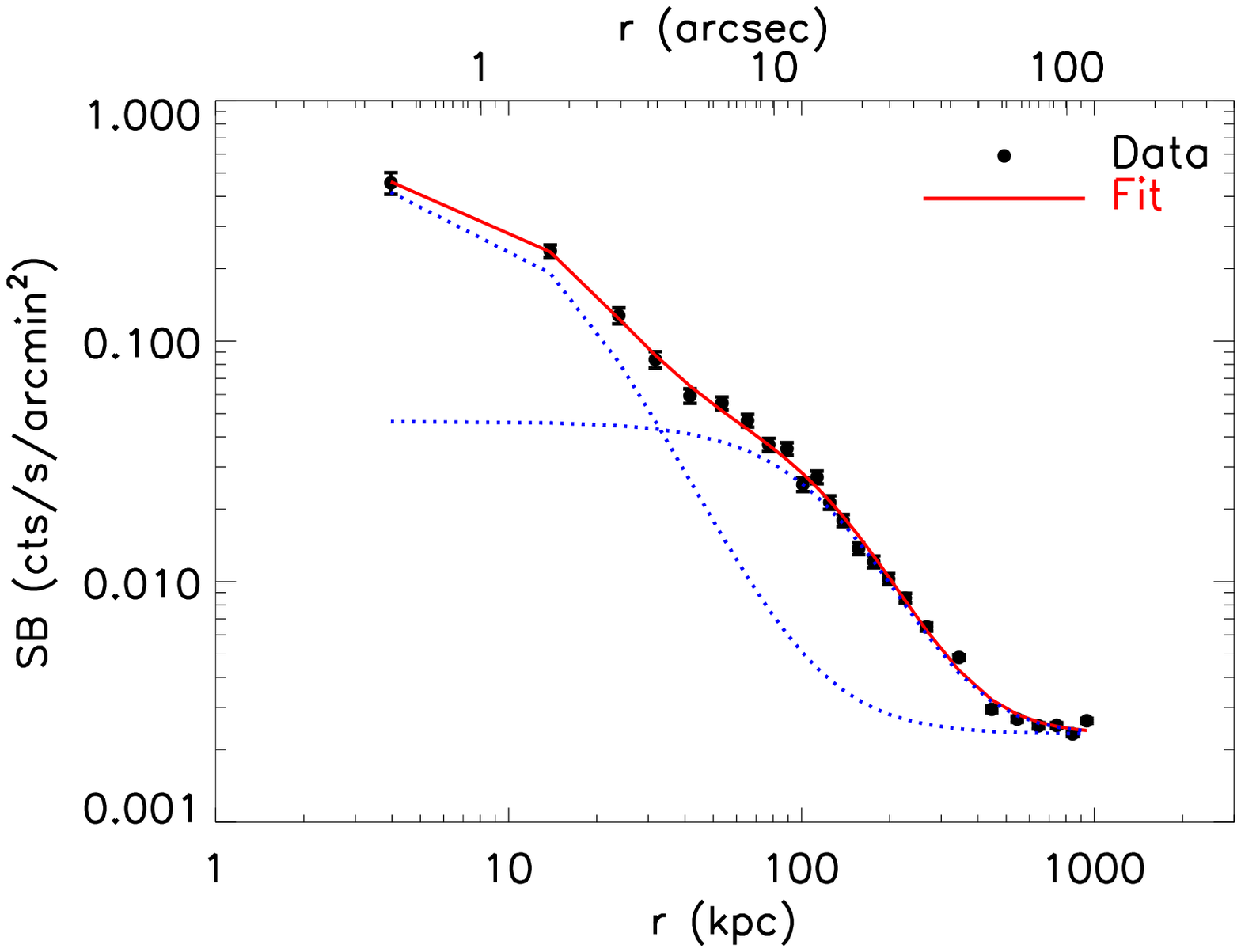}
\end{center}
\caption{Radial surface brightness profile of WAPSJ1415 (black dots). (\textit{Left}) 
Single-beta model fit and (\textit{Right}) double-beta model fit in red solid line. The two 
components of the double-$\beta$ model are shown in blue dotted lines.}
 \label{sbfit}
\end{figure*}

\begin{equation}
 S(r) = S_{0} (1+(r/r_{c})^2) ^{-3\beta+0.5} + C  
\end{equation}
\label{sbeta}

\noindent where $S_{0}$, $r_{c}$, $\beta$ and $C$ are the central surface brightness, 
core radius, slope and constant background, respectively. The fitting procedure is based on 
a Levenberg-Marquardt least-squares minimization and constrains the parameter $\beta$ 
to the range 0.4$<$ $\beta$ $<$1.0.
Local cool-core clusters require a double $\beta$-model to capture the central emission peak.
For distant clusters however, the combination of their intrinsic small angular size 
and X-ray data that usually provides only low photon statistics, make it difficult 
if not impossible to distinguish between a single- or a two-component $\beta$-model. 

We fitted the radial profile of WARPJ1415 using both the 
single- and double-$\beta$ model approximations (see Fig.~\ref{sbfit}). Given the high-quality of our data, 
we can measure a significant improvement in the fit using the double-beta model, 
with respect to the single-beta model, not only qualitatively but also statistically - see Table 2 
for a list of the fit parameters.

\subsection{Surface brightness concentration, $c_{SB}$}

In Santos et al. (\cite{joana08}), we defined the phenomenological parameter
$c_{SB}$ that quantifies the excess emission in a cluster core by measuring the 
ratio of the surface brightness (SB) within a radius of 40 kpc with respect to the average SB 
within a radius of 400 kpc: $c_{SB} =SB (r<40 kpc) / SB (r<400 kpc)$. 
This simple parameter is particularly useful when dealing with low S/N data.
The inner aperture with a physical size of 80 kpc corresponds to the 
typical size of cool-core clusters regardless of their redshift.  We stress that we use 
a physical radius instead of a scaled radius, since the cool-core phenomenon is non-gravitational 
in nature, and therefore the self-similar scaling relations are not appropriate in this case.

In Santos et al. (\cite{joana10}), we measured
$c_{SB}$=0.144$\pm$0.016 using the 90 ksec ACIS-I observation. The new,
more accurate value measured in these deep observations,
$c_{SB}$=0.150$\pm$0.007, is consistent with the previous one.

\subsection{Core luminosity excess}

Studies of nearby galaxy clusters have estimated that the ICM core contributes to about 
25\% of the total cluster luminosity (Peres et al. \cite{peres}, Best et al. \cite{best}).
The method used to compute this luminosity boost was to measure the ratio $L(r<r_{cool}) / L_{bol}$. 
We argue that this is not the most efficient way to isolate the contribution of the cool-core to the 
luminosity because a fraction of the flux in  $L(r<r_{cool})$ comes from the bulk of the cluster.

To quantify the excess luminosity due to the cool-core, we simply
computed the ratio of the flux enclosed by each of the $\beta$-model
components (core and outer part) at $r$=40 kpc (see Table 1). This is
the radius used in $c_{SB}$ and is also very close to the crossing of
the two $\beta$-model components.  Following this approach, we
estimate that the core contributes 4 times more flux relative to the
bulk of the cluster luminosity, at the cooling radius.

The cluster soft-band luminosity within $r$=400 kpc is equal to
(2.83$\pm 0.14)\times$10$^{44}$ erg~s$^{-1}$.  We extrapolate this
observed luminosity to a radius of 1 Mpc using the single $\beta$
model, and obtain $L(r<1\,\rm{Mpc})$=4.0$\times$10$^{44}$ erg~s$^{-1}$
(note that 1 Mpc is about $R_{200}$ according to our Table 1).

An accurate assessment of the excess core-luminosity has implications for the 
completeness of X-ray selected cluster samples.  In principle, in a purely flux-limited 
sample, cool-core clusters are preferentially selected with respect to non cool-core clusters with the 
same mass, thanks to their higher $L_X$.  A proper treatment of this effect will be 
relevant to properly measure the evolution of cool-cores in future X-ray surveys 
(see Santos et al. \cite{joanaproc}).

\subsection{Entropy}

\begin{figure}
\begin{center}
\hspace{-0.5cm}
\includegraphics[width=8.cm,angle=0]{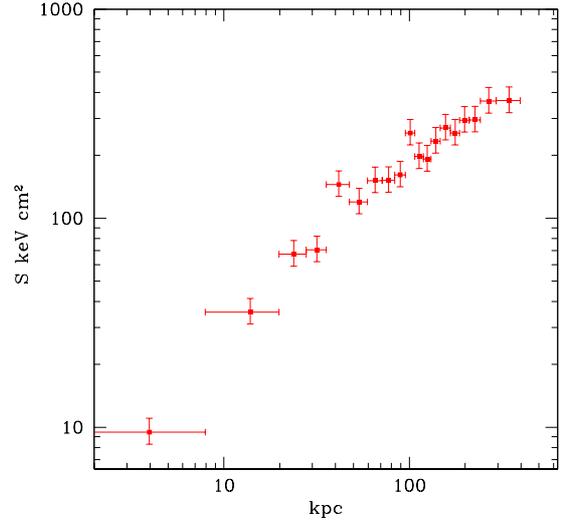}
\end{center}
\vspace{-0.8cm}
\caption{Entropy profile of WARPJ1415 computed from the numerically deprojected 
temperature and density profiles.}
 \label{entropy}
\end{figure}

Clusters are usually divided in cool-core and non cool-core on the
basis of the central value of their entropy and cooling time. However,
there are no physical reasons to expect the existence of two distinct
populations, in fact the data rather show a transition between these
two categories.  The specific entropy, $K(r)=kTn_{e}^{-2/3}$, is a
widely used quantity to describe the thermodynamical history of the
ICM. Previous studies at low-redshift have shown that cool-core
clusters have a significantly lower central entropy ($K_{c}<30$ kev
cm$^{2}$) than non cool-core clusters (Cavagnolo et
al. \cite{cavagnolo}).  Using the deprojected temperature and gas
profiles as described in Section 4, we obtain the entropy profile in
Fig.~\ref{entropy}.

In order to compare the central gas entropy of WARPJ1415
with the quoted values at low redshift, we have to keep in mind that,
since there is no defined radius to perform this measurement, the
local measurements are done in very small radii ($\le$ 1 kpc), whereas
our data do not allow us to go below $r$=8 kpc.  Thus, the central entropy
measured in the innermost bin with $r$=8 kpc,
is $K_{c}$=9.9$\pm$2.0 keV cm$^{2}$, which immediately places
WARPJ1415 in the cool-core regime.  

Note that the central entropy value obtained from the deprojected profiles refers 
to a bin centered at 4 kpc, and therefore it is nominally computed at a scale 
below 
the actual resolution of our data.  We also computed the central entropy at a radius $r=12$ kpc as in 
the first spectroscopic bin, where we actually measured the temperature in the projected data 
(see Fig.~\ref{tdeproj}, left panel).  At $r=12$ kpc the projected temperature is $kT = 3.8 \pm 0.3$, 
and this gives a value of $K_c(r=12 kpc)=20.9 \pm 2.7$ kev cm$^2$.  This value represents 
our conservative estimate of the central entropy and it is consistent with a linear relation 
$K(r) \propto r$ for $r< 40$ kpc down to $r< 10$ kpc.

\subsection{Cooling time}

The ICM central cooling time is the quantity most often used to
characterize and quantify the cool-core strength of a cluster
(e.g. Hudson et al. \cite{hudson}).  Without a heating source to
compensate radiative cooling, the ICM will radiate its thermal and
gravitational energy on a timescale $t_{cool} =p / [(\gamma -1) n_{e}
n_{H} \Lambda(T)] < 1$ Gyr (Fabian \& Nulsen \cite{fabian77}), where $p$ is the
gas pressure, $\Lambda(T)$ is the cooling function and $\gamma$ is the
ratio of specific heats of the gas.  Adopting an isobaric cooling
model for the central gas, $t_{cool}$ can be computed as:

\begin{equation}
t_{cool}(r) = \frac{2.5n_{g}T}{n_{e}^{2} \Lambda(T,Z) }   
\end{equation}
\label{tcool}

\noindent where $\Lambda(T,Z)$, $n_{g}$, $n_{e}$ and T are the cooling
function, gas number density, electron number density and temperature
respectively, with $n_{g}$=1.9$n_{e}$ (Peterson \& Fabian
\cite{peterson06}).  Local cool-core clusters are defined by a central
cooling time much lower than the Hubble time (typically $t_{cool,c}
\le$1 Gyr).

An accurate measurement of the ICM cooling time profile of distant
galaxy clusters has been unattainable up to now, due to the use of a
single ICM temperature and the average electron density instead of the
resolved profiles.  With the current data, we are able to trace the
precise cooling time profile of WARPJ1415, using the resolved
temperature, density and metal abundance profiles.  The cooling
function $\Lambda(T,Z)$, was computed using the cooling tables from
Sutherland \& Dopita (\cite{sutherland}) that account for a varying
metallicity.

By proceeding similarly to the computation of the central entropy, we
obtain a central cooling time $t_{cool}$ ranging from 0.06$\pm$0.01
Gyr at $r$ = 8 kpc to 0.23 Gyr at $r$ = 12 kpc (the central bin
of our spectroscopic analysis). This accurate measurement is a factor
14 lower than the upper limit of $t_{cool}$($r<$20 kpc)=3.4 Gyr
obtained by Santos et al. (\cite{joana10}), using the archival data
and the global cluster properties.  This result clearly demonstrates
that, unless deep, high-resolution data of high-$z$ clusters are
available, the central cooling time can only be taken as an upper
limit. 

In Table 2, we summarize the X-ray cool-core properties of WARPJ1415
using the various cool-core diagnostics described above.

\begin{table}
 \centering
\caption{X-ray cool-core estimators. The quoted temperature and metallicity values correspond to 
the projected quantities, whereas the central cooling time and central entropy values are based on deprojected quantities.}
\label{table:3}
\begin{tabular}{@{}c|c@{}}\hline
{\bf  \,\, Cool-core estimator}                         &                   \\ \hline 
  $T_{core} - T_{max} -T_{avg}$                  &  4.6 $-$ 8.0 $-$  5.7  [keV]  \\ 
$Z_{Fe,c}/Z_\odot$                                      &  3.60$^{+1.50}_{-0.85}$ \\ 
 $c_{SB}$                                                     &  0.150$\pm$0.007 \\
 t$_{cool,c}$                                                 &  0.06$\pm$0.01  [Gyr] \\
 K$_{c}$                                                       &  9.9$\pm$2.0   [keV cm$^{2}$] \\ 
\hline
\end{tabular}
\end{table}

\section{Connection between ICM core properties and the BCG}

In this paper, we also explore the connection between the radio and
optical properties of the brightest cluster galaxy with the ICM cool
core properties.  Brightest cluster galaxies hold a special place in
the history of galaxy formation and evolution.  They are the most
massive galaxies in the Universe, and are thought to have developed
through mergers as expected in a hierarchical assembly model. They are
located at the bottom of the potential well of massive clusters and are
often coincident with the X-ray peak emission of the hot gas
permeating galaxy clusters. The role of the BCG in shaping the ICM properties in the 
core as a function of cosmic epoch remains unclear, but it is expected to be an important 
factor in the ICM evolution, playing an important role in the thermodynamical 
equilibrium of the cool-core.

In the center of cool-core clusters, star formation arises as a result of the cooling
process of the ICM (Crawford et al. \cite{crawford}), with typical
star formation rates in their BCGs on the order of 1-10 $M_\odot$/yr
(O'Dea et al. \cite{odea}).  Interestingly, this connection may not be
strictly localized in the cluster center, as indicated by McDonald et
al. (\cite{mcdonald}), who found a correspondence between the spatial
location of H$\alpha$ emission regions with the X-ray morphology
(un/disturbed) of the ICM core region in nearby systems.

The radio emission of the central galaxy is related to the accretion process
onto a supermassive black hole.  The incidence rate of radio sources
in the center of cool-core clusters is at least 70\% (Best et
al. \cite{best06}, Mittal et al. \cite{mittal}; Dunn et
al. \cite{dunn08}), whereas fewer than 30\% of non-CCs host central
radio sources. Several studies have established a link between the
radio luminosity of these sources with the X-ray and optical
properties of their host clusters.  In addition, other morphological
features such as radio lobes or jets are occasionally detected.  At
redshift greater than 0.5, interactions between the central galaxy and
the ICM remain largely unexplored.

\begin{figure}
\begin{center}
\includegraphics[width=9.0cm,angle=0]{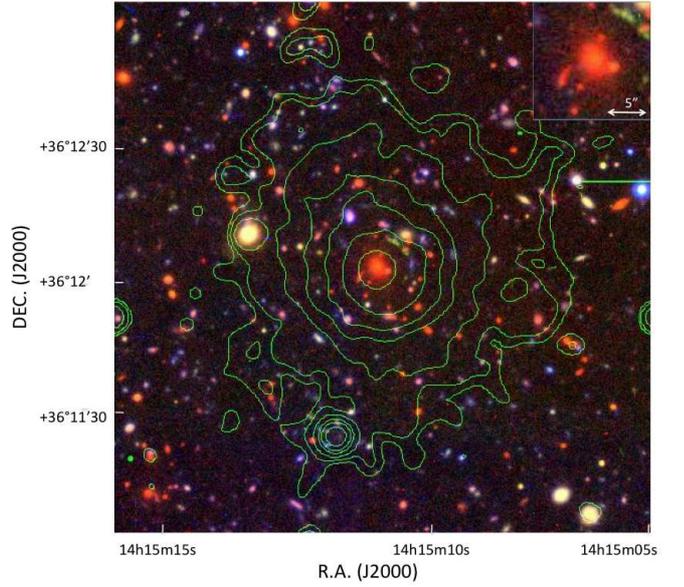}
\end{center}
\vspace{-0.5cm}
\caption{Subaru-Suprime BVR color image of WARPJ1415 covering an area with 
2$\arcmin \times$2$\arcmin$ ($\sim\! 1\times 1$ Mpc), centered at RA=14:15:11, DEC=+36:12:03. North is up 
and East is to the left. X-ray contours (with levels [3,5,10,20,30,50] $\sigma$ above the background 
and smoothed with a gaussian kernel with FWHM=5\arcsec) are overlaid in green.
The 15$\arcsec \times$15$\arcsec$ inset shows the core with a zoom factor of 2. 
}
 \label{subaru}
\end{figure}

The BCG of WARPJ1415 is very large, luminous and unusually massive,
with a reported stellar mass of 2$\times$10$^{12}$ M$\sun$ (Fritz et
al. \cite{fritz}).  The color image obtained with SUBARU-\textit{Suprime Cam} BRZ
(Fig.~\ref{subaru}) shows that the cluster central galaxy stands out as a large,
massive red galaxy, and we confirm that the BCG, the central radio
source and the peak of the X-ray emission, are spatially coincident
well within 1\arcsec. 
In this section, we analyze the radio and optical properties of the central brightest galaxy, in order to
investigate the feedback mechanism between the BCG and the ICM.


\begin{figure*}
\begin{center}
\includegraphics[width=6.cm,angle=0]{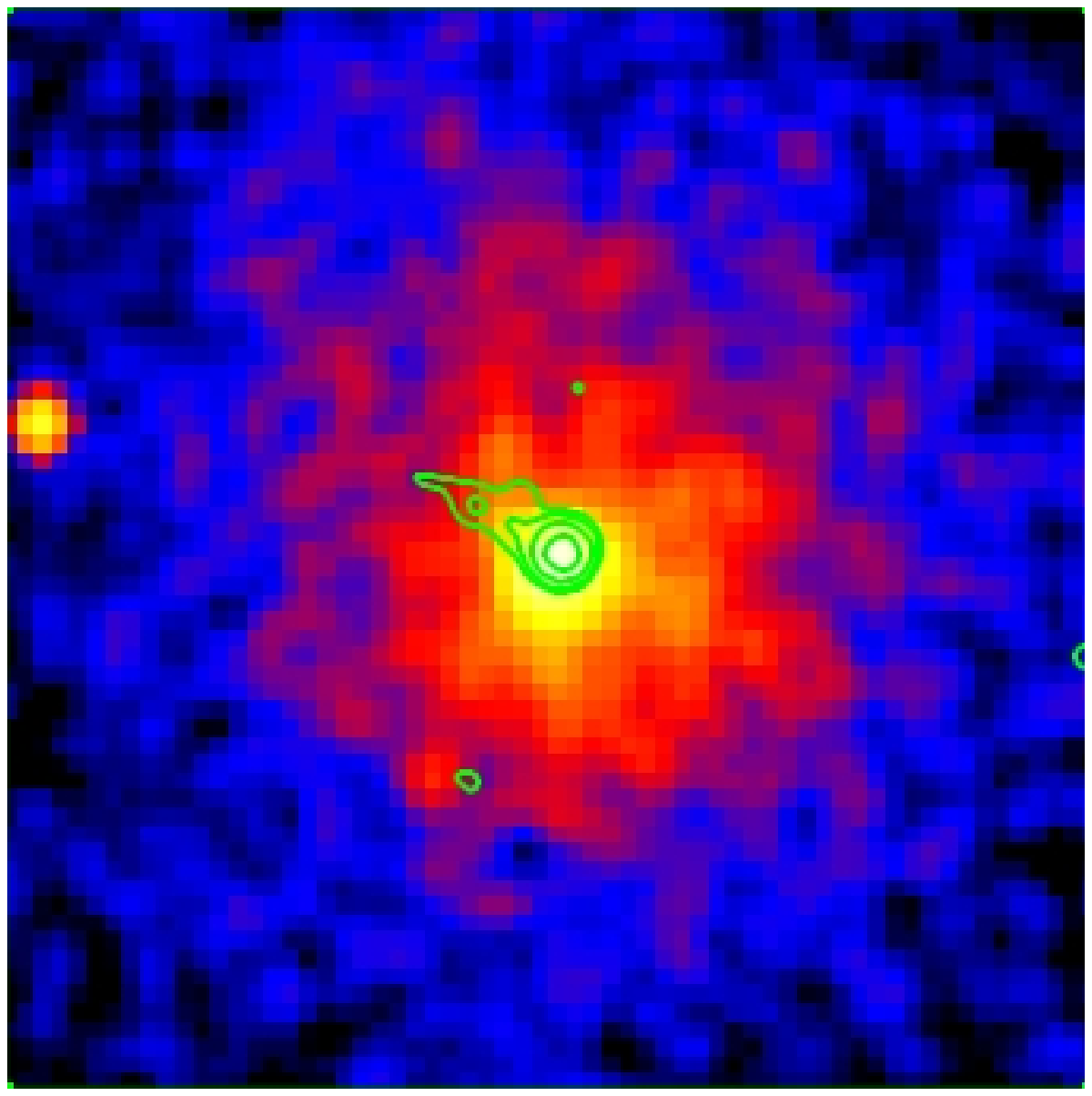}
\includegraphics[width=5.95cm,angle=0]{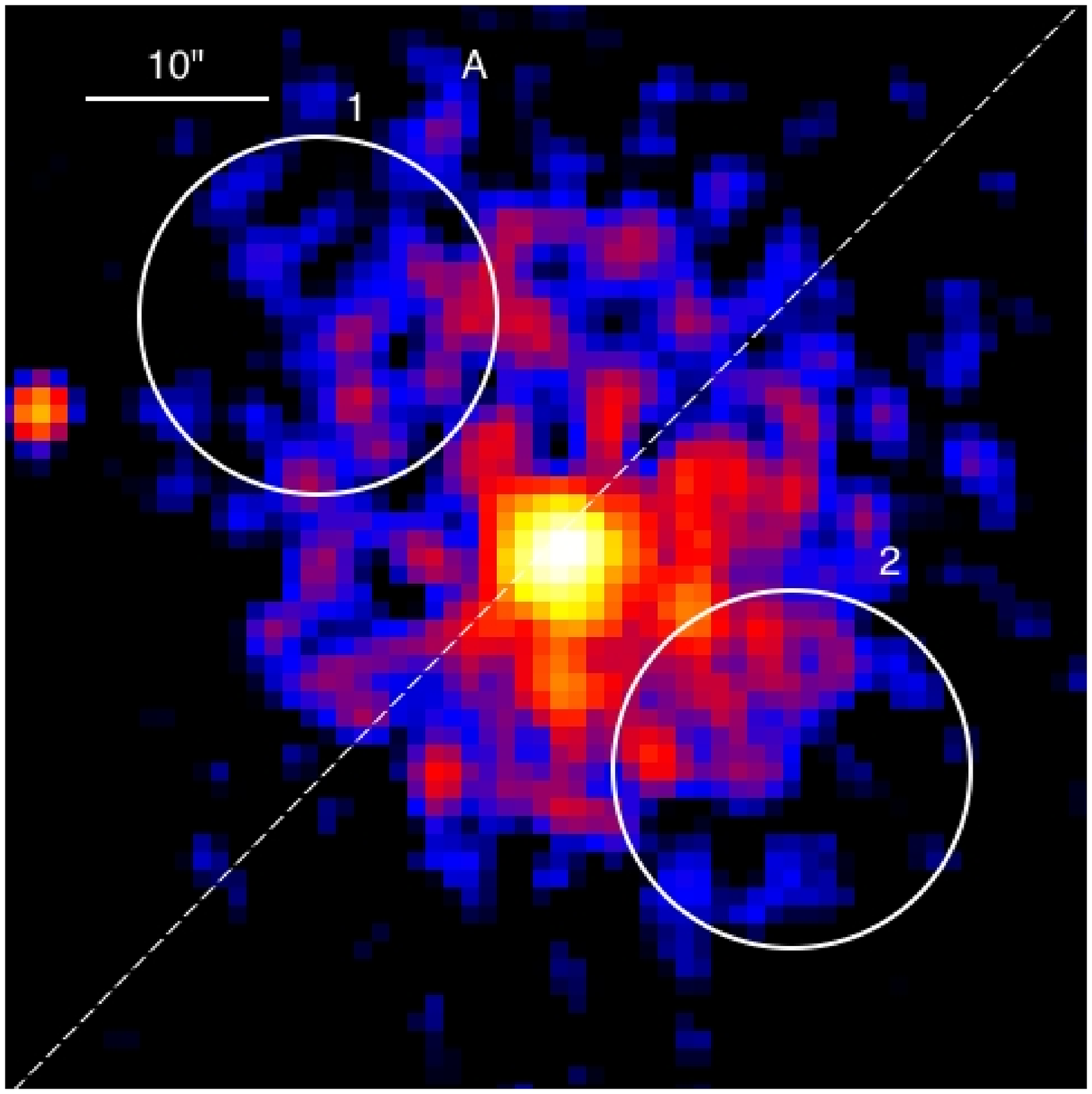}
\includegraphics[width=5.98cm,angle=0]{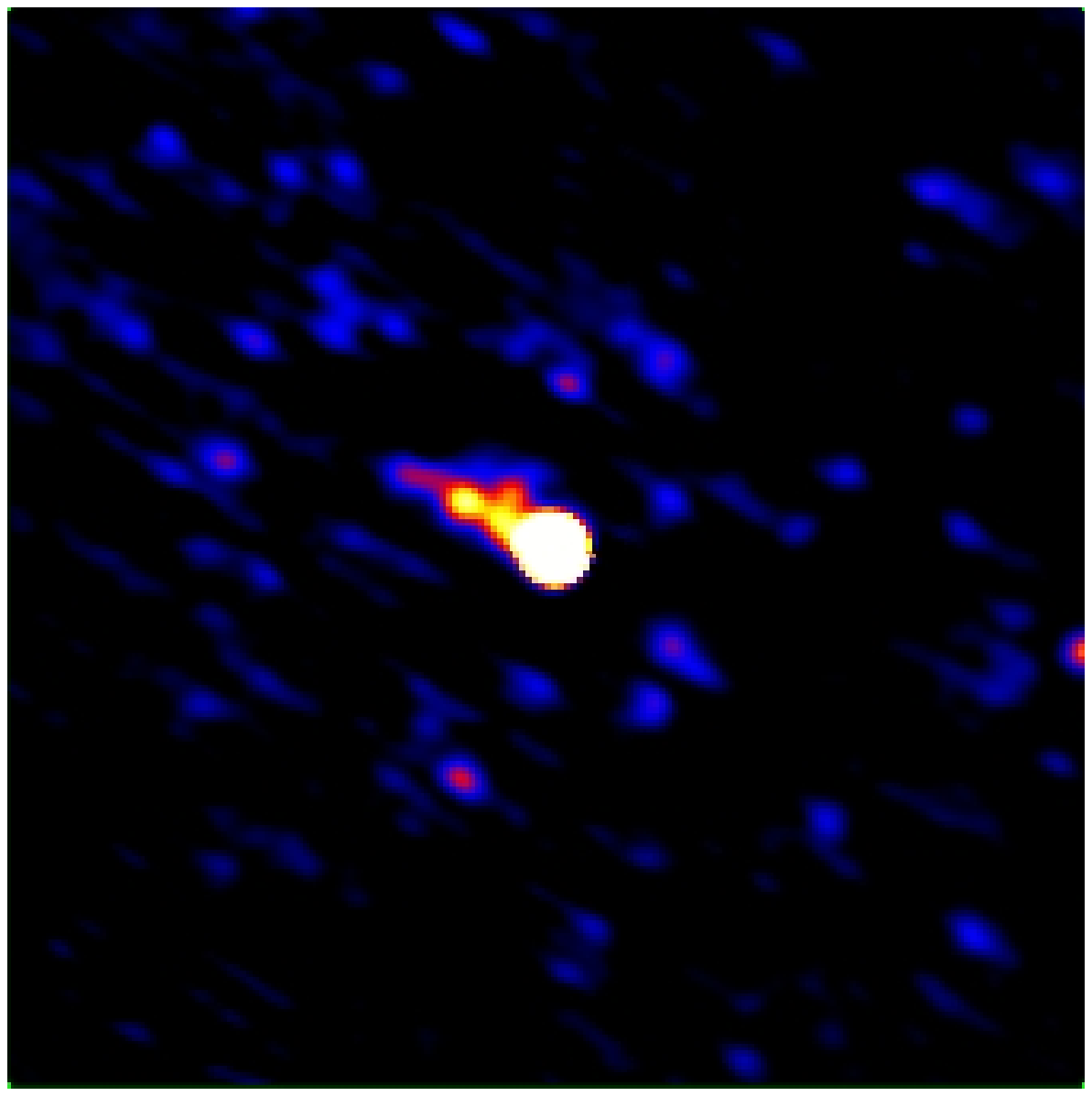}
\end{center}
\caption{(\textit{Left}) \textit{Chandra} full-band image with VLA radio
  contours overlaid in green.
  (\textit{Middle}) Residual \textit{Chandra} soft-band image after
  subtraction of the best-fit single $\beta$-model.  A lack of X-ray emission 
  is observed in sector A with respect to the average surface brightness. 
  This asymmetry is quantified by comparing the average emission in the circular 
  regions labeled 1 and 2 (see text).
  (\textit{Right}) VLA image of the radio source coincident with the cluster core.  
  A pronounced extended feature with an extent of 80 kpc is seen in the NW direction. 
  All images have a size of 1$\arcmin \times$1$\arcmin$. The \textit{Chandra} images are shown in logarithmic scale with a Gaussian smoothing of 2$\arcsec$.}
 \label{panels}
\end{figure*}

\subsection{Central radio galaxy and extended structure}

WARPJ1415 was observed with the Very Large Array (VLA) of the National
Radio Astronomy Observatory in 2002 and 2003 in the A and B
configurations at 1.4 GHz (proposal code AP439).  The combined radio map, with one hour
exposure, has a resolution of 2$\arcsec$ and a noise level of 0.016
mJy/beam.  It shows a bright radio source coincident with the X-ray
centroid of the cluster and the BCG. Thanks to our high resolution the
radio source is resolved in a bright nuclear emission (3.59$\pm$0.02
mJy) and a fainter one-sided structure in PA $\sim$ 70$^\circ$ with an
extent of $\sim$ 10$\arcsec$, corresponding to a physical size of 80
kpc at the cluster redshift (see Fig.~\ref{panels}).  The flux density of this
structure is (0.71$\pm$0.05) mJy and therefore the total flux density
of the source is 4.3 mJy. Using low resolution NVSS data we estimate a
total flux density of about 5 mJy, suggesting that in our high
resolution image all the radio source is visible.

The source radio power is in the range between Fanaroff-Riley FR I and
FR II radio galaxies, with a total radio power of 2.0$\times$10$^{25}$
W Hz$^{-1}$ ($\nu L_{\nu}$=2.8$\times$10$^{41}$ erg s$^{-1}$).
According to the low-redshift study of Sun et al. (\cite{sun}), all BCGs
with a radio AGN more luminous than 2$\times$10$^{23}$ W Hz$^{-1}$ at
1.4GHz are found to have X-ray cool-cores.  We also searched for the
location of WARPJ1415 in the BCG radio power vs. $K_{c}$ correlation
for clusters with $z <$ 0.2 presented in Cavagnolo et
al. (\cite{cavagnolo}) (Figure 2 of that paper). We find that our
cluster falls in the area under the threshold value $K_{c}<$30 keV
cm$^{2}$, and with $\nu L_{\nu} >$10$^{40}$ erg s$^{-1}$, that
characterizes nearby cool-core clusters.

The extended, asymmetric radio structure could be interpreted as a
mildly relativistic one-sided jet. However this is unlikely, since in
relatively low-power radio galaxies jets are relativistic only on a
few kpc scale, therefore with a size of $\sim$ 80 kpc we should expect
to detect radio emission also on the other side of the core. No radio
lobe seems to be present since the total radio flux density at low
resolution is not too far from our high resolution result, and
furthermore the structure is resolved also transversally. We may
interpret this radio morphology as a tail-like structure due to a
strong interaction of extended lobes with the surrounding medium.

Assuming equipartition conditions, we can estimate that in the extended emitting region
($\sim$ 40 kpc) the magnetic field is $\sim$ 6 $\mu$Gauss and the
minimum non thermal energy is $\sim$ 3 $\times$ 10$^{-12}$ erg
cm$^{-3}$.

Deeper and higher angular resolution radio data are required to confirm the 
morphology of the extended radio emission and to provide a direct evidence of a jet-ICM interaction.  
With the present data, we can only report the tantalizing hint of the feedback mechanism in action in WARPJ1415. However, the high radio power of the central AGN, and the moderate residual star formation estimated in the BCG (see next section) add further evidence in favor of a scenario in which the energy released by the radio loud AGN activity is able to stop the central cooling flow.

\subsection{X-ray cavitives?}

X-ray cavities originated by outflows of a central radio source are
occasionally detected in the ICM of local clusters (e.g. Birzan et
al. \cite{birzan04}, Fabian et al. \cite{fabian06}).  The detection of
such bubbles is challenging even at low redshift due to their low
surface brightness contrast and small sizes that range from $<$1 kpc
to $\sim$40 kpc (larger cavities are very rare).  Detection rates are
of the order of 20-25\% in the \textit{Chandra} archive (Rafferty et
al. 2008) and in the flux-limited B55 cluster sample (Dunn et
al. \cite{dunn05}). Nearby strong cool-core clusters have a much
higher incidence rate, reaching 70\%.
 
Even though we have a deep, high-resolution observation, distant
clusters have a small angular size which makes the detection of X-ray
cavities very difficult.  We searched for X-ray bubbles and/or surface
brightness asymmetries that could be associated with the extended
radio emission. In order to do this we explored several techniques to
improve the image contrast, namely, by subtracting both single and
double $\beta$-models to the original \textit{Chandra} soft-band
image, and by creating an unsharp-masked image.  In the residual
images it was difficult to see a convincing indication of a cavity.
However, there is a clear asymmetry in the residual image obtained by
subtracting the single-$\beta$ model (see Sect. 5.1) to the X-ray
data, with a significant lack of X-ray emission in the upper half of
the SB map (labeled sector A in Fig.~\ref{panels}), in the direction of the
extended radio emission (green contours). 
We quantified this asymmetry by measuring the number of net photons in two circular 
regions with a radius equal to 10\arcsec in the exposure-corrected \textit{Chandra} image 
(see circle 1 and 2 in the middle panel of Figure 9).  The two numbers should be consistent with each
other if no asymmetry is present.  Instead, we measure $192\pm 14$ photons in circle 2 and 
$153 \pm 12$ photons in circle 1, showing that region 1 (the one in the direction of the jet) is
 25\% less luminous with respect to region 2 at a 2-$\sigma$ confidence level.

\subsection{Equivalent width of the [OII] emission line}

Optical emission lines (H$\alpha$ at 6563 \AA) are a relatively
dust-independent measure of recent star formation. In alternative, the
[OII] ($\lambda$ 3727) emission line is a good proxy for H$\alpha$.
Brightest cluster galaxies are usually red, early-type galaxies,
undergoing passive evolution.  Hence, star-formation is seldom found
in these central, massive, cluster galaxies.  In spite of this, high
star formation rates have been found in BCGs of local clusters with
central cooling times shorter than $\sim$ 0.5 Gyr (Rafferty et
al. \cite{rafferty08}; Cavagnolo et al. \cite{cavagnolo}), most likely
caused by a residual cooling flow.

The brightest central galaxy of WARPJ1415 was observed with the Gemini
Multi-Object Spectrograph (GMOS-N, Hook et al. \cite{hook}) in 2003
(PI Ebeling).  We reduced the archival data with IRAF procedures and
obtained the final spectrum, with a scale of 1\AA/per pixel and a resolution $R \approx$ 1000, as the
result of the median stacking of 10 spectra with a total exposure time
of 5.3 h and a S/N$\sim$5. We expect no major slit loss since the
average seeing is 0.7$\arcsec$ and the width of the slit is
1$\arcsec$.

A significant, broad [OII] emission line is found at the cluster
redshift.  The broadening of optical emission lines originated by warm
ionized gas has been often observed in the BCGs of cool-core clusters and is
likely a Doppler effect caused by motions in the center of the galaxy
(Heckman et al. \cite{heckman}).

We measured the equivalent width (EW) of the [OII] line defined by 
\begin{equation}
EW= \int \frac {F_{c}-F_{\lambda}} { F_{c}} d\lambda
\end{equation}
\noindent where $F_{c}$ is the continuum flux and $F_{\lambda}$ is the
flux of the emission line.  The rest-frame EW of the [OII] line in the
central galaxy of WARP1415 is -25$\pm$3 \AA. 
Our result goes against the interpretation by Samuele et
al. (\cite{samuele}) of a strong decrease of star formation activity
with redshift, based on the EW of the [OII] line in 77 BCGs selected
from the 160 SD survey.  In that work, BCGs with an EW stronger than
-15 \AA~are not present.  As mentioned in Santos et
al. (\cite{joana10}), we suggest that cool-core clusters might be
under represented in the 400 SD sample (and hence its subset, the
160 SD sample), which might also explain the lack of strong [OII]
lines in the clusters' BCGs.

\subsection{Star-formation rate of the BCG}

We estimate the intrinsic [OII] line luminosity assuming
\textit{E(B-V)}=0.3, using the following relation from Kewley et
al. (\cite{kewley}):

\begin{equation}
L\rm{[OII]_{i}} (erg\, s^{-1}) = L [OII]_{o} \times10^{0.572}  
\end{equation}

\noindent where $L$[OII]$_{o}$ is the observed [OII] luminosity.
The [OII] line falls in optical i-band at the cluster redshift, therefore we used the high-resolution \textit{HST/ACS} 
F775W ($i'$-band) archival observations to determine the luminosity of the [OII] line. We scaled the 
observed spectrum to the F775W magnitude of the BCG measured within a radius of 0.75$\arcsec$, 
in the interval encompassed by the F775W filter. 

To derive the galaxy star formation rate, we applied the modified Kennicutt law 
corrected for reddening as presented in Kewley et al. (\cite{kewley}), considering 
a mass range of 0.1-100 $M_\odot$ for a Salpeter (Salpeter \cite{salpeter}) initial mass function:

\begin{equation}
{\rm SFR \, [OII]}\, (M_\odot \, yr^{-1}) = (6.58\pm1.65)\times10^{-42} L\rm{[OII]}. 
\end{equation}

\noindent Depending on the assumptions made on the amount of dust, we
provide a range of SFR.  In a dust-free scenario, we estimate SFR=2.2
$M_\odot/yr$, instead, if we assume a reddening correction of
\textit{E(B-V)}=0.3, a value typically used in star-forming galaxies
(e.g. Lemaux et al. \cite{lemaux}), we obtain SFR=8.3 $M_\odot \,
yr^{-1}$.

Alternatively, the [OII] line emission in red-sequence galaxies of high-redshift
clusters ($z\sim$0.9) may be caused by an AGN (Lemaux et
al. \cite{lemaux}). Although we cannot entirely rule out emission from
an AGN contributing to the estimated star formation rate in WARPJ1415,
the fact that we do not detect an X-ray point source coincident with
the BCG supports the assumption that the [OII] emission can be entirely 
ascribed to star formation processes associated to the residual cooling flow in the core.
In a forthcoming paper, we will further explore
the properties of the BCG and the cluster galaxy population, using a
rich optical-IR dataset, including the photometry used here from
HST/ACS, SUBARU/\textit{Suprime} in addition to \textit{Spitzer}-IRAC.
The use of infrared data will allows us to firmly disentangle a
possible AGN contamination to the SF rate diagnostics.

\begin{figure}
\begin{center}
\hspace{-0.5cm}
\includegraphics[width=7.8cm,angle=0]{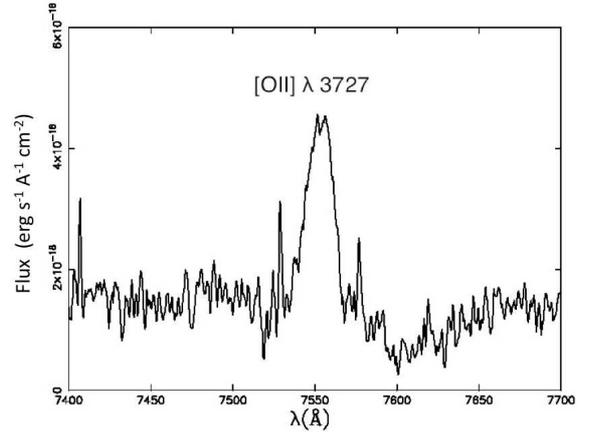}
\end{center}
\caption{Observed optical GMOS spectrum of the BCG of WARPJ1415 showing a prominent, 
broad [OII] emission line. }
 \label{lowz}
\end{figure} 
 
 \subsection{Fe excess and the ICM metal enrichment process}

 The interplay between the BCGs and their host clusters were first
 studied by Edge \& Stewart (\cite{edge}).  They found a correlation between
 the optical luminosity of the BCG with the X-ray luminosity and hot
 gas temperature of its host cluster. This work was expanded by De
 Grandi et al. (\cite{degrandi04}), where the excess iron mass in the
 core region of the ICM has been shown to correlate with the optical
 and NIR luminosity of the BCG, suggesting a co-evolution between the
 stellar content of the galaxy and the ICM metal abundance.

We measured a total Fe mass $M_{Fe} = 6.2^{+2.5}_{?1.9}\times 10^{9} M_\odot$ 
(within $R_{200}$) using the best-fit abundance profile (see Section 4).  For the purpose 
of our study, it is more interesting to measure the Fe mass excess, M$_{Fe}^{exc}$,
following the definition of De Grandi et al. (\cite{degrandi04}). This is simply the residual iron mass 
within 200 kpc, obtained by subtracting the almost constant abundance value measured at 
$R > 200$ kpc ($0.25 Z_\odot$ in units of Asplund et al. \cite{asplund}).  We measured 
 M$_{Fe}^{exc}$=1.8$_{-0.5}^{+0.7} \times 10^{9 }M_\odot$ which corresponds to about 30\% of 
the total Fe mass. This number is larger than the average value of 10\% found in nearby 
cool-core clusters studied in De Grandi et al. (\cite{degrandi04}), but the difference is not significant due 
to the large scatter in the observed local values.  In fact, we find that our cluster lies in good
agreement with the correlation between the iron mass excess measured in local cool-cores and 
their temperature.

To probe the connection between the optical properties of the central galaxy and the metal enrichment of the ICM, 
De Grandi et al. (\cite{degrandi04}) found a correlation between the absolute optical magnitude of their local 
BCGs computed with the Rc-band 
with the excess iron mass. 
Assuming that the iron excess is originated entirely by the BCG, this relation may imply that the efficiency 
of the metal transport mechanisms  from the galaxy to the ICM may be approximately the same in all clusters.

In order to check whether WARPJ1415 is consistent with this correlation found in local cool-cores, we measured the 
aperture magnitude of the BCG 
using the \textit{Suprime} Rc-band. 
The k-correction was computed with a  SWIRE elliptical template of 5 Gyr, generated with the GRASIL code 
(Silva et al. \cite{silva}). 
We obtain an absolute magnitude of -23.8 mag, therefore WARPJ1415 falls in the expected
$M_{optical} - M_{Fe}^{exc}$ relation for local cool-core clusters (see Fig. 10 of De Grandi et al. \cite{degrandi04}).
In the assumption that the excess iron mass is entirely due to Type Ia SNe and stellar mass loss in the BCG,
we can immediately place an upper limit to the time-scale needed to build up the iron peak.
We can safely assume that the bulk of the stars in the BCG formed at about $z\gtrsim$2 (Renzini \cite{renzini06}). 
We also consider $z$ = 3 as upper limit to the epoch when most of the stars in BCG have formed, as indicated by 
the observation of major star-formation episodes in massive spheroids at $z \ge$ 2 (Daddi et al. \cite{daddi}) or stellar 
population modeling of high-z BCGs (e.g. Rosati et al. \cite{rosati09}). 
This corresponds to a lookback time of 10.3 Gyr (11.5 Gyr at $z$=3) in the adopted cosmology.
Since WARPJ1415 lies at a lookback time of nearly 7.9 Gyr, the build up of its metal peak must have
happened on a timescale of $\sim$2.4 Gyr with an upper limit of 3.6 Gyr. Our results point towards
times scales shorter than previously claimed ($>$5 Gyr see B\"ohringer et al. \cite{boehringer}) to develop a supra-solar 
iron abundance peak. This would imply a SNe Ia rate larger than expected at $z>1$ in the BCG, an 
evidence which should be compared with the recent first estimate of Type Ia SN rate in high-z clusters 
(Barbary et al. \cite{barbary}).
In a forthcoming paper (Tozzi et al. in prep.) we will investigate the far reaching consequences these data
have on chemical enrichment models,
specifically how one can constrain the time scale of the iron production rate through SNe Ia, 
and the relative contributions of SNe Ia, stellar mass loss and SNII.

\section{Conclusions}

In this paper we presented a unique \textit{Chandra} observation of a
distant galaxy cluster.  These data enabled a spatially resolved
analysis of the ICM, with the main goal of studying the properties of
the cluster cool-core, and trace a connection with the central galaxy,
using optical and radio data. We measured the following X-ray
properties:

\begin{itemize}
\item a significant temperature decrease towards the center:
  $T_{c}$=4.6$\pm$0.4 keV, is 2.2 keV lower than the global
  temperature, $T$=6.8 keV and 3.4 keV lower than the maximum
  temperature at $r$=90 kpc, $T$=8.0 keV;
\item a surprisingly pronounced metallicity peak, $Z_{Fe,c}$=
  3.60$^{+1.50}_{-0.85} Z_\odot$ with a corresponding excess iron mass
  M$_{Fe}^{exc}$=1.8$_{-0.5}^{+0.7} \times 10^9 M_\odot$;
\item a central surface brightness excess modeled by a double-$\beta$ model
and quantified by $c_{SB}=0.150\pm 0.007$.  The contribution of the cool-core 
to the X-ray luminosity within $r=40$ kpc is 4 times the value obtained by 
extrapolating the surface brightness profile from outside the cool-core towards the center;
\item the central cooling time, $t_{cool,c}$=0.06 (0.23) Gyr and the
  central entropy, $K_{c}$=9.9 (20.9) keV cm$^{2}$ at $r=8$ (12) kpc
  (the values in parentheses correspond to a more conservative
  estimate);
\item using the measured temperature and density profiles, the cluster
  total X-ray mass, under the assumption of hydrostatic equilibrium, is
  $M_{200}$=3.0$_{-0.4}^{+0.6}\times$10$^{14}$M$_\odot$.
\end{itemize}

Using VLA high-resolution data we detected a source coincident with
the BCG with a radio luminosity $L_{1.4 GHz}$=2.0$\times$10$^{25}$ W
Hz$^{-1}$. A faint, one-sided structure with an extent of 80 kpc is
seen in the north-west direction, where a significant lack of X-ray
emission was found.  Furthermore, the analysis of optical spectroscopy
of the central galaxy shows a broad [OII] emission line, with a
moderate to strong equivalent width of -25 \AA, corresponding to an associated star
formation rate in the range [2.3-8.3] $M_\odot/yr$.

We were also able to trace a connection between the radio and optical
properties of the central galaxy with the X-ray cool-core.
By comparing the correlations among the radio luminosity, SFR and $K_{c}$ 
measured with low-redshift clusters with our results, we confirm the same 
feedback mechanism at work in the core of WARPJ1415.

We were also able to investigate the connection between the radio and the optical properties 
of the central galaxy with those of the ICM properties in the core.  In particular, the ratio 
between radio luminosity, $K_{c}$ magnitude and the estimated SFR in the BCG of
WARPJ1415 is consistent with what is found in low-redshift clusters. These findings 
suggest that the feedback mechanism at work in the core of WARPJ1415 is of the same 
kind and intensity of that observed in local clusters.

The prominent Fe peak
indicates that the metal enrichment mechanisms by type Ia supernovae (SN Ia)
and star formation in the BCG happened on a short timescale (given the
lookback time of 7.8 Gyr), and/or that the transport processes that
drive away the metals to the outskirts (e.g. galactic winds) were not
efficient to smear out the Fe excess.

The analysis of the \textit{Chandra} data shows that WARPJ1415 at
$z$=1 has all the classical features that characterize nearby
cool-core clusters.  These observations enabled the most detailed
analysis of the ICM of a $z \ge$1 cluster and our results highlight the
importance of deep, high-resolution data to adequately characterize
distant clusters. We were able to obtain a noticeable improvement on
the previous X-ray characterization and our results set strong
constraints on cluster evolution models.  

Our results are a first step towards understanding a possible
cosmological evolution of the central feedback in galaxy clusters, and
refining AGN feedback prescriptions in galaxy evolution models at
higher redshifts. To this aim, clearly a sizable representative sample
of distant clusters with similar data quality, i.e. S/N
and angular resolution, is needed. This is a task which only
\textit{Chandra} can currently achieve, albeit with a substantial
investment of time, whereas it is within comfortable reach of
 next generation wide field X-ray telescopes (e.g., the Wide Field 
 X-ray Telescope, Giacconi et al. \cite{giacconi}).
 In Santos et al. (\cite{joanaproc}), we studied how the measure of the parameter $c_{SB}$ varies 
with redshift and angular resolution using simulated clusters.  Based on this study we predict that we will 
be able to properly evaluate $c_{SB}$ for WARPJ1415 with an angular resolution of 5\arcsec (Half Energy Width) 
or better.  We conclude that only a combination of a large effective area and a large field of view, when 
coupled with a constant angular resolution of the order of 5\arcsec, can provide a significant breakthrough on the 
study of cool-core clusters over a wide redshift range up to $z$=1.5, both in terms of statistics and
data quality.  With current X-ray facilities, we expect that only few cases can be studied with an accuracy 
comparable to that of WARPJ1415.

\begin{acknowledgements}
We thank Italo Balestra for reducing the XMM-Newton data of WARPJ1415, 
and Sabrina De Grandi for useful comments on the metal enrichment of the ICM.
We also thank the Chandra team for their assistance.
This work was carried out with Chandra Observation Award Number 12800510 in GO 12.
We acknowledge support under grants ASI-INAF  I/088/06/0 e ASI-INAF I/009/10/0.
      PT acknowledges support under the grant INFN PD51.
\end{acknowledgements}

\bibliographystyle{aa}

\end{document}